\documentclass[aps,prd,twocolumn,showpacs,floatfix,superscriptaddress,
preprintnumbers,nofootinbib]{revtex4}

\usepackage{graphicx,amssymb,url,subcaption,xcolor}

\newcommand{\be}{\begin{equation}}
\newcommand{\ee}{\end{equation}}
\newcommand{\ba}{\begin{eqnarray}}
\newcommand{\ea}{\end{eqnarray}}

\def\lsim{\raise0.3ex\hbox{$\;<$\kern-0.75em\raise-1.1ex\hbox{$\sim\;$}}}
\def\gsim{\raise0.3ex\hbox{$\;>$\kern-0.75em\raise-1.1ex\hbox{$\sim\;$}}}
\def\eps{\varepsilon\def\theta{\vartheta}}
\def\ap{\approx}
\def\calE{\mathcal E}

\begin{document}

\title{High energy electromagnetic cascades in  
extragalactic space: physics and features}

\author{V.~Berezinsky}
 \affiliation{INFN, Gran Sasso Science Institute 
viale F.Crispi 7, 67100 L'Aquila, Italy and 
Laboratori Nazionali del Gran Sasso, Assergi (AQ), 67010,  Italy}

\author{O.~Kalashev}
\affiliation{Institute for Nuclear Research of the Russian Academy of
 Sciences,  Moscow 117312, Russia}

\date{\today}

\begin{abstract}
Using the analytic modeling of the electromagnetic cascades compared
with  more precise numerical simulations we describe the physical properties 
of electromagnetic cascades developing in the universe on CMB and EBL
background radiations. A cascade is initiated by very high energy photon
or electron and the remnant photons at large distance have two-component
energy spectrum, $\propto E^{-2}$ ($\propto E^{-1.9}$ in numerical 
simulations) produced at cascade multiplication stage, and 
$\propto E^{-3/2}$ from Inverse Compton electron cooling at low energies.  
The most noticeable property 
of the cascade spectrum in analytic modeling is 'strong universality', 
which includes the standard energy spectrum and the energy density
of the cascade $\omega_{\rm cas}$ as its only numerical parameter.
Using numerical simulations of the cascade spectrum and comparing it 
with recent Fermi LAT spectrum we obtained the upper limit on  
$\omega_{\rm cas}$ stronger than in previous works. The new feature of
the analysis is  ''$E_{\max}$ rule''. We investigate the dependence of  
$\omega_{\rm cas}$ on the distribution of sources, distinguishing two 
cases of universality: the strong and weak ones.  

\end{abstract}


\maketitle

%%%%%%%%%%%%%%%%%%%%%%%%%%%%%%%%%%%%%%%%%%%%%%%%%%%%%%%%%%%%%%%%%%%%%%
\section{Introduction}
%%%%%%%%%%%%%%%%%%%%%%%%%%%%%%%%%%%%%%%%%%%%%%%%%%%%%%%%%%%%%%%%%%%%%%
Very high energy extragalactic electron or photon colliding with 
low-energy background  photons $\gamma_t$ (CMB and EBL) produces 
electromagnetic cascade  due to reactions 
$\gamma + \gamma_t \to e^-+e^+$ (pair production, PP) and 
$e+ \gamma_t \to \gamma'+e$ (Inverse Compton 
scattering, IC). Part of the cascade energy can be taken away by 
synchrotron radiation if magnetic field is strong enough. Existence 
of such cascading was understood soon after discovery of Cosmic 
Microwave Background (CMB) radiation \cite{CMB} and of 
Greisen-Zatsepin-Kuzmin cutoff \cite{GZK} (oral remark by  
I.L.~Rozental at Soviet-Union Cosmic Ray Conference, 
by personal recollection of one of us, VB, and independently by  
S.~Hayakawa according to recollection of K.~Sato). Since that time 
e-m cascading in extragalactic space received many various
applications.       

One of the earliest applications was rigorous upper limit on cosmogenic 
neutrino flux, which was  proposed by Berezinsky and Zatsepin in 1969 
\cite{BZ1969}. The upper limit on this flux as was first obtained in  
\cite{BS1975} follows from observation that production of 
neutrinos and initial cascading particles ($e$ and $\gamma$) takes
place from decay of the same particle, $\Delta^+$ resonance, produced 
in $p\gamma_{\rm cmb}$ interaction, and thus energy density of produced neutrinos 
and cascade particles are characterised by ratio $1/3$. A remarkable byproduct 
of this work was a good agreement of analytically calculated cascade spectrum 
with diffuse extragalactic gamma-ray spectrum as presented finally by EGRET 
collaboration in 1998 \cite{EGRET} in the range from 30~MeV to 130~GeV. 
The exponent of EGRET spectrum $\alpha=2.1 \pm 0.03$ agrees quite well
with predicted in \cite{BS1975} as $\alpha=2.0$. The diffuse gamma-ray 
spectrum measured recently by Fermi-LAT \cite{FermiLAT,Fermi2014} showed 
much worse agreement with predicted cascade spectrum, in particular 
the spectrum exponent is found  $\alpha=2.3$. The allowed energy
density of the cascade radiation is found \cite{BGKO2011}  
\be
\omega_{\rm cas}= 5.8 \times 10^{-7}~ {\rm eV/cm}^3,
\label{eq:omega2011}
\ee
and it puts the severe limit on allowed flux of cosmogenic neutrinos 
and extragalactic protons in Ultra High Energy  Cosmic Rays (UHECR) 
\cite{BGKO2011},\cite{Ahlers2010},\cite{GKS2012}. 

The new Fermi data \cite{FermiLAT} probably contradict also the more general  
early hypothesis that extragalactic background gamma-ray flux observed 
by EGRET \cite{EGRET} and Fermi-LAT \cite{FermiLAT} is fully produced 
in $p\gamma$ interaction of extragalactic CRs followed by e-m cascading. 
This hypothesis, put forward in early 70s e.g. in  \cite{Prilutsky} 
and \cite{SWW}, had the impact especially on interpretation of
the EGRET data, but more complicated Fermi-LAT spectrum and in
particular discovery of discrete sources in the early background spectrum 
seriously questioned this interpretation. 

Electromagnetic cascading strongly affected gamma-astronomy of 
discrete sources. The first work in this field was calculation
by Gould and Schreder \cite{GouldSchreder} of absorption of gamma-rays 
with energies above 100~TeV on CMB radiation in the Universe. Cascading 
of the absorbed photons was understood soon and in 1970 the absorption
of UHE photons on optical, radio-radiation and in magnetic fields 
were included in the calculations, together with similar calculations
for electrons \cite{Berez1970}.   

The new step was done in the work by Aharanian, Coppi and V\"olk 1994
\cite{Felix}. Before this work magnetic field in cascading process was
taken into account for energy losses of electrons and for absorption
of photons in very strong magnetic field. In the work \cite{Felix} 
the authors noticed importance of deflection of the cascade electrons 
in magnetic fields. In absence of magnetic field cascade particles 
propagate from a source in the same direction as parent photon. If 
extragalactic magnetic field nearby the source is large enough, the 
low-energy cascade 
$e^+e^-$-particles can be deflected  from the direction of initial  
cascading photon and produce (by IC radiation) isotropic low-energy 
$E_\gamma < 1$~TeV component named by the authors the ``halo component''. 
As the  sources the AGN and in particular blazars are considered. In 
terms of presently estimated EBL radiation~\cite{Kneiske:2003tx}, 
\cite{Kneiske:2010pt} the considered model looks  as follows.  An initial 
photon with energy $E_{\gamma0} \sim 10$~TeV is absorbed on EBL with 
the mean absorption length $\ell_\gamma \sim 100$~ Mpc, producing 
electron and positron with energy $E_e \sim 5$~TeV each. 
Electron/positron is deflected in extragalactic magnetic field 
producing then in IC scattering on a CMB photon the recoil photon with 
energy 
\be
E_\gamma^{\rm IC} \sim (4/3) \gamma_e^2 \eps_{\rm cmb} \sim 100~{\rm GeV}     
\label{eq:E_IC}
\ee
where $\gamma_e=E_e/m_e$ is the Lorentz-factor of electron. Thus 
a typical energy of halo radiation is $E \sim 100$~GeV and the size of
halo is $r_h \sim~$a few Mpc. 

A very exiting application of cascading was started by the work 
of  Neronov and Semikoz \cite{NeronovSemikoz} who indicated a
possibility to search for very weak seeds of magnetic fields in 
the Universe.

Creation of the seeds with extremely weak magnetic fields is 
a necessary part of explanation of observed magnetic fields 
which can reach the tremendous values up to $10^{13}$~G  
deduced for pulsars. The strong magnetic fields can be 
produced very fast due to collapse of the objects with 
weak magnetic field and fast increasing of magnetic fields 
due to dynamo mechanism. The problem is how the objects with 
very weak magnetic fields, the seeds, were produced. Two kind of
the seeds are in principle known: the cosmological and astrophysical
ones (for the reviews see e.g. \cite{Kronberg}, \cite{GrassoRubinstein},   
\cite{KulsrudZweibel} and latest review with many references 
\cite{DurrerNeronov}. Magnetic seeds of astrophysical origin include
historically the first model ``Biermann battery'' \cite{LBiermann},
\cite{MestelMoss} and recently many models based on plasma
instabilities e.g. \cite{MedvedevLoeb} and also on different models 
for Population III stars, e.g. \cite{MiniatiBell}.

To measure magnetic fields in seeds Neronov and Semikoz 
\cite{NeronovSemikoz} suggested to observe cascading propagation  
of TeV gamma-rays from a source through a void with very weak
magnetic fields. Secondary positrons and electrons are weakly
deflected in magnetic fields producing thus extended emission of 
IC gamma-rays, i.e. gamma-ray halo described above. Decrease of the
size of this halo with energy of emitted gamma-rays allows to measure 
the strength of magnetic field $B$ in a range  
$10^{-16}~{\rm G} \lsim B \lsim 10^{-12}$~G \cite{Elyiv2009}.

At present from observation of cascading radiation of TeV gamma-ray
sources (blazars) it became possible to put the lower limit  
on extragalactic magnetic field \cite{NeronovVovk2010}, for the review 
and other references see \cite{DurrerNeronov}.  

Following the references cited above and observational data of
spectra of TeV blazars obtained by Fermi, one may explain an 
appearance of lower limit on extragalactic magnetic field 
in the following way. 

Consider first the case $B=0$ and gamma-radiation from a blazar 
with primary energies higher than 1~TeV directed along the jet 
to an observer. These photons are absorbed leaving behind the 
cascade radiation with low-energy spectra $\propto E^{-2}$ below   
cutoff and $E^{-1.5}$ at smaller energies (see \cite{BS1975} and 
section \ref{sec:physics}). These predicted low-energy spectra 
exceed the Fermi observation at $E \sim 1 - 100$~GeV and most natural 
assumption is suppression of these fluxes due to magnetic field, 
which deflects $e^+e^-$ pairs from direction to an observer. The 
MC simulations from~\cite{NeronovVovk2010,Taylor:2011bn} and other calculations cited 
in~\cite{DurrerNeronov} result in the lower bound $B > 10^{-17}$~G 
for extragalactic magnetic fields in voids.  All these limits depend 
on size of coherence length of magnetic field $\lambda$. 

Apart from lower limits obtained using cascading from different blazars
there is one case of the positively measured magnetic field applying
quite different method suggested by T.~Vachaspati  and his
collaborators \cite{TashiroVachaspati},\cite{ChenVachaspati} and
references therein.
This case is relevant for cascading in helical 
magnetic field. Such field scatters $e^+e^-$ and  provides non-zero 
correlator between positions of three cascading gamma-quanta produced 
by IC radiation of electrons and positrons. The data of Fermi from blazars 
are used for the analysis. The helical magnetic field $B \sim 10^{-14}$~G 
is found on 10~Mpc scale. This helical magnetic field can be produced
in early universe at $t \sim 1$~ns.  

Extragalactic magnetic fields in other structures, like filaments and
galaxy clusters are larger by many orders of magnitudes reaching 
$\mu$G level in clusters. A reader can find more wider and detailed
discussion of  extragalactic  magnetic fields in the review~\cite{DurrerNeronov}. Above we limited ourselves by several issues
connected with cascading.

This paper is organized in the following way.
In the section II ``Cascade physics and analytic calculations''  we
describe the cascade physics and obtain analytical solutions of
cascade equations. 

Our basic model in the section \ref{sec:physics} is a static universe 
filled by background radiations with dichromatic spectrum of 
photons with energy $\epsilon_{\rm cmb} =6.3\times 10^{-4}$~eV for 
CMB and $\epsilon_{\rm ebl} =0.68$~eV for EBL radiation, the only 
free parameter in this model. Cascade is initiated by very high 
energy electron or photon with energy $E_s$ and develops due
to pair-production $\gamma +\gamma_t \to e^++e^-$  and 
and inverse-compton (IC) $e +\gamma_t \to e' +\gamma'$ scattering on 
background target photons $\gamma_t$. At large enough time the spectrum 
of remnant photons obtains {\em universal} form (independent from 
initial energy $E_s$ assuming $E_s$ is high enough), which has
universal characteristic energies: energy of the spectrum cutoff    
$$
\calE_{\gamma}^{ebl}=\frac{m_e^2}{\epsilon_{\rm ebl}}=
3.9\times 10^{11}~{\rm eV},
$$
and energy of spectrum steepening
$$
\calE_X= \frac{1}{3}\calE_{\gamma}^{\rm ebl}\frac{\epsilon_{\rm cmb}}
{\epsilon_{\rm ebl}}= 1.2\times 10^8~{\rm eV}
$$
The spectrum of remnant photons is given by 
$n_\gamma(E) \propto E^{-3/2}$ at $E \leq \calE_X$, 
$n_\gamma(E) \propto E^{-2}$ at $\calE_X\leq E\leq\calE_\gamma$,
$n_\gamma(E)=0$ at $E\geq\calE_\gamma$,

The spectrum $n_\gamma(E) \propto E^{-3/2}$ at $E \leq \calE_X$ is 
robust. The spectrum at $\calE_X\leq E\leq\calE_\gamma$ is approximate: 
numerical simulation give the exponent $\gamma=1.9$ instead of 2.0. 
The cutoff is given in rough approximation. 

In subsection \ref{sec:universal} we demonstrate that the spectra 
obtained in static universe using the analytic dichromatic model have 
the property of universality, which we will later call the 
strong universality.
The main feature of universal spectrum is its fixed shape, independent 
of initial energy $E_s$ and distance to the source. The universality 
is broken for nearby sources. 

In subsection~\ref{sec:comparison} the analytic  universal spectrum is 
compared with numerical simulations for the cascades initiated at
red-shift $z$. For $z=0.15$ (r=626 Mpc) agreement for dichromatic 
model with $\epsilon_{\rm ebl}=0.68$~eV is good, as it is good for
larger $z$ but with choice of different values of $\epsilon_{\rm ebl}$. 
The case of small distance (low $z$) needs the different treatment.

In subsection \ref{sec:cascadeCMB} we consider physically and technically 
interesting cases of CMB-only radiation. This case is important at
large $z$ when density of CMB photons strongly dominates, and for nearby 
sources when absorption on EBL is small or absent. Technically the case 
of CMB-only is interesting because it automatically produces dichromatic
effect: the role of EBL photons is played by the photons with energy 
$\tilde{\eps}_{\rm cmb}$ from high-energy tail of Planckian
distribution whose density is enough to absorb HE cascade 
photon with energy E of interest at considered distance $r$.  

In this approximation we calculated the cascade spectrum from nearby
sources r= 1~Mpc,~~ 8.5~Mpc,~~ 85~Mpc and 200~Mpc and compared them 
with numerical simulations also in assumption of CMB-only background
radiation. The most noticeable difference with numerical simulation  
is observed at smallest distance  r= 1~Mpc because in realistic
calculations the cascades arrive to an observer as under-developed 
with spectrum  $\propto E^{-1.47}$. The calculated parameters of the
cascades are shown in Table \ref{table1} of subsection 
\ref{sec:nearby}. The new element of calculations is low-energy
suppression of the spectrum shown by $\calE_{\rm lec}^\gamma$ 
in Table \ref{table1}.

In the section~\ref{sec:simul} we discuss the two numerical simulation 
techniques for calculation of cascade spectrum, namely solution 
of one-dimension transport equation and Monte Carlo simulation. The 
former method is much faster and allows precise calculation of propagated 
spectra in the case when cascade deflections are not important, while the 
latter method is good for investigation of the effects of magnetic
field, e.g. in a problem of isotropization of the cascades 
emitted by point sources. We compare our numerically-calculation spectra 
obtained with two techniques with each other and  with independent 
numerical simulations.

In the section~\ref{sec:universality} the cases of strong and weak
universality are introduced and discussed. In section \ref{sec:omega} 
the corresponding spectra  are calculated and compared with that measured 
by Fermi LAT. This comparison allows to obtain upper limits on energy 
density of the cascade radiation $\omega_{\rm cas}$, shown in 
Fig.~\ref{omega} as function of production redshift. The limits are
stronger for generations of the cascades at small redshifts. 

In section \ref{sec:conclusion} we give a summary of the paper with 
the main conclusions. This section is written in autonomous way and  
is destined for a reader who is interested in the main results and
wants to bypass the technical details.

We use in the paper the following abbreviations: 

UHECR for Ultra High Energy Cosmic Rays,~~ CMB for Cosmic Microwave
Background,~~ EBL for Extragalactic Background Light,~~ EGB for 
Extragalactic Gamma-Ray Background from Fermi-LAT data,~~ 
IGRB for Isotropic Gamma Ray Background from Fermi-LAT data,~~
IC for Inverse Compton, PP for Pair Production,~~ MC for Monte Carlo,
~~ IGMF for Inter Galactic Magnetic Field,
~~  MAGIC for Major Atmospheric Gamma Imaging Cherenkov Telescope, 
H.E.S.S for High Energy Stereoscopic System,

%%%%%%%%%%%%%%%%%%%%%%%%%%%%%%%%%%%%%%%%%%%%%%%%%%%%%%%%%%
\section{Cascade physics and analytic calculations}  
\label{sec:physics}
We develop here a simplified model for the cascade which allows 
us to obtain an approximate spectrum of sterile (remnant) photons 
left behind the cascade multiplication.  We
perform  the approximate calculations of e-m cascade from a burst 
of radiation in a form of very high energy electrons or photons at 
very large distance from an observer. One may think of a single 
electron or photon with very high energy $E_s \gsim 10^{15}$~eV
or a number of such particles. We use the {\em dichromatic
spectrum} of background radiation which 
consists of photons  with energy $\epsilon_{\rm cmb}$ and 
$\epsilon_{\rm ebl}$, analogues of cosmic microwave radiation (CMB)
and extragalactic background light (EBL) with fixed energies of order 
of $6.3\times 10^{-4}$~eV and $\sim 1$~eV, respectively. We assume 
$\epsilon_{\rm ebl} >> \epsilon_{\rm cmb}$ for energies and  
$n_{\rm cmb} >> n_{\rm ebl}$ for space densities. A cascade is 
initiated by a single electron or photon with very high energy $E_s$  
and proceeds through pair-production (PP), 
$\gamma + \gamma_{\rm bckgr} \to e^{-}+e^{+}$, and inverse compton (IC) 
scattering, $e + \gamma_{\rm bckgr} \to e'+\gamma'$, on low-energy 
background photons (CMB or EBL).  We assume low magnetic field which 
does not influence the cascade development due to synchrotron
radiation, and our calculations will be mostly concerned with the mean 
diffuse photon flux which is not affected by deflections of the cascade 
electrons and positrons in the magnetic field.

We consider first the flat static universe and calculate the cascade 
spectrum from a point-like burst of radiation at a large distance from 
an observer. The remnant cascade photons at large distances become the 
cascade-sterile and we refer to spectrum of these photons as universal. 

To compare the calculated spectrum with numerical simulations 
we consider further expanding universe and a burst of high energy 
radiation at a point with redshift $z$. In this case the spectrum of 
sterile photons remaining from e-m cascade undergoes the redshift.   
%%%%%%%%%%%%%%%%%%%%%%%%%%%%%%%%%%%%%%%%%%%%%%%%%%%%%%%%%%%%%%%%%%%%%%
\subsection{Universal spectrum in analytic calculations}
\label{sec:universal}

 In this subsection we consider the flat static universe and e-m cascade 
from a point-like burst of very high energy photons or electrons 
at distances from an observer to be large enough for remnant photons to
become sterile.

\begin{figure*}[h]
\begin{center}
\includegraphics[width=2\columnwidth]{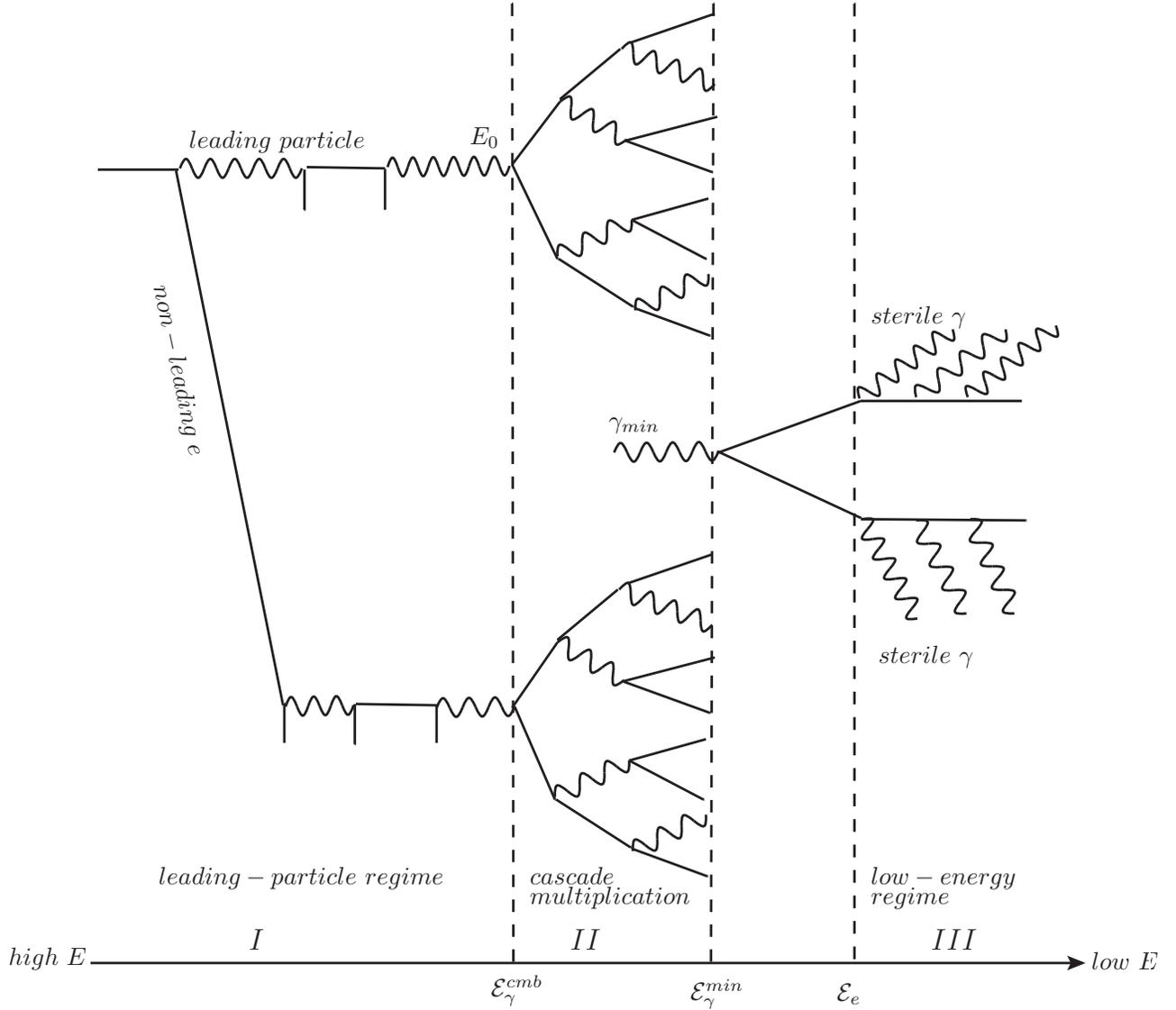}
\caption{Qualitative picture of the cascade development in 
static universe for mono-energetic energies of background  
photons $\epsilon_{\rm cmb}$ and $\epsilon_{\rm ebl}$ with 
assumption $\epsilon_{\rm ebl} \gg \epsilon_{\rm cmb}$ and 
$n_{\rm cmb} \gg n_{\rm ebl}$. The energies of cascade particles 
$\calE_{\gamma}^{cmb}$, $\calE_{\gamma}^{\min}$ and $\calE_e$ mark 
three regions of the cascade development: (I) HE leading-particle regime
when a leading particle losses very small energy in a collision
with background photons, (II) fast cascade multiplication, with
comparable fraction of energy obtained by each of two produced particles 
and (III) regime of production of sterile photons by the electrons with 
energy $E \leq \calE_e$  (see text for the details.)}
\label{fig:scheme}
\end{center}
\end{figure*}

The criterion of high energy (HE) and low energy (LE) for a cascade 
particle with energy $E$ is given with help dimensionless parameter $x_t$ 
\be
x_t= E\eps_t/m_e^2,
\label{eq:x}
\ee
where $\eps_t$ is energy of a target photon (t=cmb or ebl).
$x_t\gg 1$ and $x_t \ll 1$ characterizes HE and LE regimes, respectively.
In HE regime, for both PP and IC, a cascade particle
propagates as a leading particle $\gamma \to e \to \gamma \to e$ 
(see Fig.~\ref{fig:scheme}, leading-particle regime) loosing in each 
collision (PP and IC) fraction of energy \cite{VB1970}:
\be
f \ap 1 / [\ln (2E \eps /m_e^2)] .
\label{eq:fraction}
\ee

The turning point occurs when a leading particle approaches  
$x \gsim 1$ , and enters {\em multiplication regime} II in 
Fig.~\ref{fig:scheme}.  
\vspace{2mm}

In case of expanding universe we assume that development of the cascade 
occurs during time  $\tau_{cas}(E)$ much shorter than the Hubble time  
$H^{-1}(z_b)$, where $\tau_{cas}(E) \sim 1/(\sigma(E) c n_{\rm bckgr})$ 
and $\sigma(E)$ is cross-section for HE photon or electron. The cascade 
development continues until the cascade photons reach, due to
multiplication, the threshold of pair-production in the process 
$\gamma+\gamma_{\rm ebl} \to e^++e^-$. After this moment the remnant 
cascade photons  loose energy only by redshift.

Because of inequality $n_{\rm cmb} \gg n_{\rm ebl}$ the cascade
development proceeds in two stages. At the first one a cascade  
develops in collisions with CMB photons only: 
$\gamma + \gamma_{\rm cmb} \to e^{-}+e^{+}$ and  
$e + \gamma_{\rm cmb} \to e'+\gamma'$. At the second stage the remnant
photons from HE part of the distribution are absorbed on EBL radiation  
$\gamma + \gamma_{\rm ebl} \to e^{-}+e^{+}$, and then the produced 
$e^+$ and $e^-$ are scattered on more numerous CMB photons: 
$e + \gamma_{\rm cmb} \to e'+\gamma'$. Because of this, calculations of 
the cascade on CMB background only, has the physical importance. This
stage becomes particularly significant at $z_b \gg 1$ when the
EBL radiation is absent and a cascade develops only on CMB. At smaller 
readshifts EBL radiation appears, and the cascade enters its second stage. 
\vspace{2mm}

The remnant photon spectrum is characterized by the following benchmark
energies: the minimum energy of absorbed photon $\calE_{\gamma}^{\min}$,
to which we refer to as {\em min-photon}, the minimum energy of absorbed 
photons for CMB background only $\calE_{\gamma}^{cmb}$, minimum energy
of the cascade electron/positron $\calE_e=\calE_{\gamma}^{\min}/2$ 
({\em min-electron}) and the energy $\calE_X$ of photon ({\em X-photon}) 
produced in IC by min-electron. 
These energies are listed below in Eq.(\ref{eq:benchmark}) 
together with their numerical values estimated for values 
$\epsilon_{\rm cmb} = 6.3\times 10^{-4}$~eV and   
$\epsilon_{\rm ebl} = 0.68 $~eV (see discussion below). 

\begin{eqnarray}
\calE_{\gamma}^{\min}=\calE_{\gamma}^{ebl}=\frac{m_e^2}{\epsilon_{\rm ebl}}=
3.9\times 10^{11}~{\rm eV}\nonumber\\
\calE_{\gamma}^{cmb}=\frac{m_e^2}{\epsilon_{\rm cmb}}=4.1\times 10^{14}~
{\rm eV}\label{eq:benchmark}\\
\calE_e=\frac{1}{2}\calE_{\gamma}^{\min}=1.95 \times 10^{11}~{\rm eV}\nonumber\\
\calE_X= \frac{1}{3}\calE_{\gamma}^{\min}\frac{\epsilon_{\rm cmb}}
{\epsilon_{\rm ebl}}= 1.2\times 10^8~{\rm eV}\nonumber
\end{eqnarray}
These energies appear in the cascade development  
illustrated by Fig.~\ref{fig:scheme}.

At highest energies $x \gg 1$ the cascade develops in leading particle 
regime I, $\gamma \to e \to \gamma \to e$, with a small fraction of energy
(\ref{eq:fraction}) lost in every collision. The non-leading (nl) particle 
in this process is always electron (or positron). However, as long as 
$E_e^{\rm nl} \eps /m_e^2 \gg 1$ non-leading electron propagates as 
a leading particle and transfers its energy to cascade multiplication 
regime II (see Fig.~\ref{fig:scheme}, leading-particle regime I). As 
numerical calculations show, the total energy injected into regime II
approximately equals to initial energy $E_s$, mainly because the non-leading
electrons can enter the multiplication regime (see Fig.~\ref{fig:scheme}).

The leading-particle regime finishes at $\calE_\gamma^{\rm cmb}$, when 
$e^+e^-$ pairs are not produced on CMB photons and approximately at this 
energy cascade multiplication regime II starts. Pair-production there 
occurs on EBL radiation while IC proceeds mainly on more numerous CMB 
photons.  The minimum energy 
$\calE_{\gamma}^{\min}=3.9\times 10^{11}$~eV of photons, 
which are able to produce $e^+e^-$ pairs on high-energy EBL target  
photons, mark the end of the cascade multiplication and beginning  of 
low-energy region III where IC on CMB photons, 
$e + \gamma_{\rm cmb} \to e'+\gamma'$, dominates. The produced photons
are cascade-sterile and, in consideration of this subsection,
propagate not loosing the energy adiabatically. 
  
The energy spectrum of these photons is easy to calculate from two 
relations: $dn_\gamma =dE_e/E_\gamma$ and 
$E_\gamma \approx (E_e/m_e)^2 \epsilon_{\rm cmb}$, valid for 
low-energy regime III. One obtains:
\be 
dn_\gamma/dE_\gamma \propto E_\gamma^{-3/2}
\label{eq:LEspectrum}
\ee
Eq.~(\ref{eq:LEspectrum}) gives the robust prediction for low-energy
asymptotics of the cascade spectrum. 

There is one feature common for all stages of the cascade I - III:
due to strong dominance of density of CMB photons, 
$n_{\rm cmb} \gg n_{\rm ebl}$, IC scattering is alsways dominated by CMB 
photons. 

For this regime in region III and most part of region II the
relation given by equation (\ref{eq:ICcmb})  below is valid:
\be 
E'_{\gamma}=\frac{4}{3}\gamma_e^2 \epsilon_{\rm cmb},
\label{eq:ICcmb}
\ee
 where $\gamma_e = E_e/m_e$ is the electron Lorentz-factor and 
$E'_{\gamma}$ is the average energy of cmb photon after scattering off 
the electron. This equation was obtained in 
\cite{Blum1970} and \cite{VB1970} and it is valid at energy of 
target photons $\eps_r \ll m_e$ in the system where the electron 
is at rest.

\begin{figure}[t]
\begin{center}%
\includegraphics[angle=0,width=0.5\textwidth]{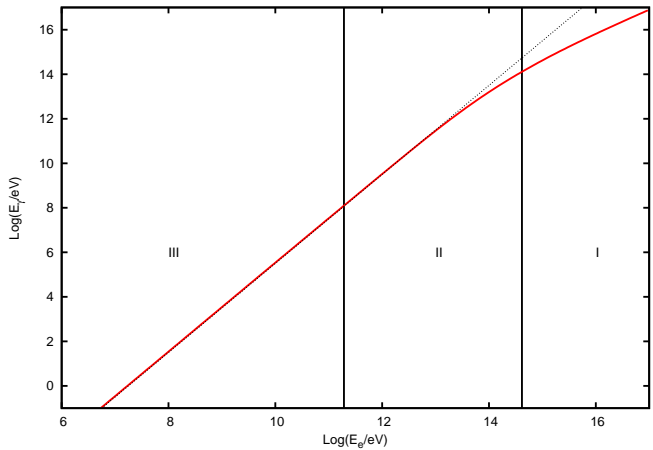}
\end{center}
\caption{Comparison of the exact calculation of mean energy of the recoil 
CMB photon $E'_\gamma$ in IC scattering  (shown by red curve) and its 
asymptotic dependence according to Eq.~(\ref{eq:ICcmb}) (black dotted line). }
\label{fig:EsecGamma}
\end{figure}

The essential feature of Eq.(\ref{eq:ICcmb}), 
 $E'_\gamma \sim \gamma_e^2 \epsilon_{\rm cmb}$ can be easily obtained 
from the Lorentz transformations. Indeed, consider  an electron with 
Lorentz factor $\gamma_e$ colliding with CMB photon $\epsilon_{\rm cmb}$.
The energy of this target photon in the system of electron at rest, 
$\epsilon_r \sim \gamma_e \epsilon_{\rm cmb}$ is assumed to be much
 smaller than $m_e$. After scattering such photon does not change its 
energy $\epsilon'_r=\epsilon_r$, and in laboratory system it is  typically 
boosted by another Lorentz factor $\gamma_e$: 
$\epsilon' \sim \gamma_e \epsilon'_r \sim \gamma_e^2\epsilon_{\rm cmb}$.
 In Fig.~\ref{fig:EsecGamma} we compare $E'_\gamma$ from 
Eq.~(\ref{eq:ICcmb})
(black dotted line) with exact calculation (red line) both given as a function 
of electron energy  $E_e$ plotted at abscissa. One may see that
maximum energy $E'_{\gamma}$ from Eq.~(\ref{eq:ICcmb}) is valid up to 
$3\times 10^{13}$~eV. In fact in all cases of applications below 
the maximum energy of photon spectra is provided by absorption on EBL
radiation and thus maximum energy is given by   
$\calE_{\gamma}^{ebl}=\frac{m_e^2}{\epsilon_{\rm ebl}}=
3.9\times 10^{11}~{\rm eV}$, much below $3\times 10^{13}$~eV allowed by
exact calculations. The only almost exceptional case 
$\calE_X = 2.0 \times 10^{13}$~eV is given by nearby sources 
(see Table\ref{table1} in Section \ref{sec:nearby}). Probably 
too flat $\propto E^{-1.7}$ part of the spectrum between $\calE_X$ 
and  $\calE_\gamma$ is a reflection of more flat red curve in 
Fig.~\ref{fig:EsecGamma} above $E'{_\gamma}= 3\times 10^{13}$~eV.  

%%%%%%fig:EsecGamma}

It is important to note that instead of following the time-dependent 
cascade development, we consider a history of a cascade in terms of 
{\em particle generations}. Particles, electrons and photons, 
can reach given generation $\nu$ at different times. The total number of 
particles in one generation is $N_{\rm tot}=2^{\nu}$ and relation between 
number of electrons $N_e$ and photons $N_\gamma$ in the same
generation with large $\nu$ is approximately $N_e \approx 2N_{\gamma}$ 
(in case a cascade starts by electron  this relation is given by 
$N_e= 2N_{\gamma} + (-1)^\nu$ and in case by photon 
$N_e=2N_\gamma - 2(-1)^\nu$).

For calculation of the 
cascade energy spectrum  we introduce the quantity $q(E)$, as a number 
of cascade particles passing through energy $E$ during the whole time of 
cascade propagation. For electrons and photons we use notation $q_e(E)$ 
and $q_{\gamma}(E)$, respectively. Assuming that total energy in the
cascade is conserved, one can use the equality of primary-particle energy  
$E_s$ and the total energy $E\times q(E)$ which flows through energy $E$
during all cascade history. Taking into account that $N_e \approx 2N_\gamma$
in each generation of the cascade in the multiplication region II, one 
obtains $q_e(E)=(2/3)E_s/E$ and $q_\gamma(E)=(1/3)E_s/E$ (for different 
way to prove $q(E) \propto (1/E)$ see \cite{BS1975} and \cite{book}).
For low-energy regime III we have  $q_e(E)=const$ for electrons and  
$q_\gamma(E)$ increasing with energy due to low-energy tail of photons 
produced by electrons.

The basic equation for the number of the cascade photons 
$n_\gamma(E)$ reads
\be 
dn_\gamma(E_\gamma) = q_e(E_e) dE_e/E_\gamma .
\label{eq:n_gamma}
\ee
In low-energy regime III we can use additionally to Eq.~(\ref{eq:n_gamma})
$q_e(E_e)=q_0$ at $E_e \leq \calE_e$, and $E_\gamma \propto E_e^2$ for
IC photon production on CMB photons. 
It results in $dn_\gamma/dE_\gamma \propto E_\gamma^{-3/2}$ at 
$E_\gamma \leq \calE_X$ in agreement with Eq.~(\ref{eq:LEspectrum}).  

In multiplication regime II we use additionally to basic equation  
(\ref{eq:n_gamma}), $q_e(E_e) \propto 1/E_e$ and $E_\gamma \propto E_e^2$
We obtain thus $dn_\gamma/dE_\gamma \propto E_\gamma^{-2}$, valid 
in the interval $\calE_X \leq E_\gamma \leq \calE_\gamma^{\min}$.
At $E_\gamma \geq \calE_\gamma^{\min}$ all remnant photons are
absorbed and $dn_\gamma/dE_\gamma=0$.

Thus we finally obtain for the spectrum of remnant photons in terms of
the total number of particles $n_\gamma$:
\be
n_\gamma(E_\gamma)=\left\{ \begin{array}{ll}
(K/\calE_X)(E_\gamma/\calE_X)^{-3/2} ~&{\rm at}~~ E_\gamma \leq \calE_X \\ 
(K/\calE_X)(E_\gamma/\calE_X)^{-2} ~&{\rm at}~~ \calE_X\leq E_\gamma \leq\calE_\gamma\\
0               ~&{\rm at}~~  E_\gamma > \calE_\gamma
\end{array}
\right. 
\label{eq:gamma-spectrum}  
\ee
where absorption energy 
$\calE_\gamma=\calE_\gamma^{\min}=m_e^2/\eps_{\rm ebl}=3.9\times 10^{11}$~eV,
is minimum energy of absorbed photon (min-photon), and transition energy
$\calE_X=(4/3)(\calE_e/m_e)^2\eps_{\rm cmb}=1.2\times 10^8$~eV is the
energy of IC photon radiated by min-electron produced in absorption
of min-photon. Everywhere below we use $\calE_\gamma$ and $\calE_X$ 
as generic notation for cutoff energy and energy of transition
between $E^{-3/2}$ and $E^{-2}$ regimes, respectively. The background
radiation (cmb or ebl) can be indicated as indices, when it is
needed. The normalizing coefficient K can be found from
conservation of energy in the cascade, with the total energy equals 
to energy $E_s$ of primary electron or photon.  
\be
K=\frac{E_s}{\calE_X(2+\ln \calE_\gamma/\calE_X)}.    
\label{eq:Knorm}
\ee
We will refer to spectrum (\ref{eq:gamma-spectrum}) with normalization
(\ref{eq:Knorm}) as {\em universal spectrum}: its shape is fixed
independently on initial energy $E_s$ or even injection spectrum
$Q(E_s)$, if initial energy is sufficiently high, larger than scale 
energy $E_0$, which can be taken as $\calE_\gamma^{\rm cmb}$ 
from Eq.~(\ref{eq:benchmark}). 

The initial energy $E_s$
(or energy density $\omega_{\rm cas}$ for diffuse flux) changes only 
total normalization coefficient $K$ from Eq.~(\ref{eq:Knorm}). 
Thus the shape of the cascade spectrum does not depend on the
injection spectrum and propagation time (spectrum is frozen at 
the remnant-photons stage and forgets about its production stage).
This universality will be referred to as 'strong universality'.

The universal spectrum given by Eq.~(\ref{eq:gamma-spectrum}) with 
normalization (\ref{eq:Knorm}) has been obtained above in the simple 
model for static flat universe, but in fact it is valid for a
wide class of different models, e.g. for realistic expanding universe 
at fixed redshift, if cascade develops at time $\tau$ shorter than the 
Hubble time $H^{-1}(z)$. In many cases below, Eqs~(\ref{eq:gamma-spectrum}) 
and (\ref{eq:Knorm}) are valid, though with numerical values of 
$\calE_\gamma$, $\calE_X$ and $K$ different from that given  above. 

The universal spectrum of remnant photons left behind 
cascading of particles on dichromatic background photons with energies 
$\eps_{\rm cmb}=6.3\times 10^{-4}$~eV and $\eps_{\rm ebl}=0.68$~eV 
in flat universe at large distance from an observer is characterized
by the following spectral features:
it has  a flat energy spectrum $\propto E_\gamma^{-3/2}$ from the lowest
energies and up to $\calE_X = 1.2\times 10^8$~eV; it becomes more
steep $\propto E^{-2}$  at higher energies, followed by a sharp cutoff  at
$\calE_\gamma=m_e^2/\eps_{\rm ebl}=3.9\times 10^{11}$~eV due to
absorption on EBL photons. The most reliable predictions of this simple
model are low energy spectral shape $\propto E^{-3/2}$ (robust) and 
cascade-multiplication one $\propto E^{-2}$ (probably approximate), while
prediction of sharp cutoff at $\calE_\gamma$ is caused by assumption of 
large distance to the  source and monochromatic spectrum of EBL. The 
sharpness of the spectral feature at $\calE_X$ with its numerical value 
is artifact of dichromatic model of background radiation. In fact at 
both energies $\calE_X$ and $\calE_\gamma$ there must be the transition 
regions. 

The low-energy component of spectrum (\ref{eq:gamma-spectrum}) 
$\propto E^{-1.5}$ is a signature of low-energy IC scattering 
(\ref{eq:ICcmb}), while the component $\propto E^{-2}$ is a signature
of the cascade multiplication (region II in Fig.~\ref{fig:scheme}
up to cutoff at energy  $\calE_{\gamma}^{ebl}= 3.9\times 10^{11}~{\rm eV}$   
).

Eq.~(\ref{eq:gamma-spectrum}) gives the total number of photons 
from a point-like source. In particular in case of one primary UHE 
photon/electron $n_\gamma(E_\gamma)dE_\gamma$ gives the total 
number of the cascade photons with energy $E_\gamma$ observed 
at any large distance $r$ from a source. Since the both characteristic
energies $\calE_\gamma$ and $\calE_X$ do not depend on distance 
and energy $E_s$, Eq.~(\ref{eq:gamma-spectrum}) presents the same 
universal spectrum $n_\gamma(E_\gamma)$ for any initial energy 
$E_s$ being larger than energy $\calE_\gamma$. Eq.~(\ref{eq:gamma-spectrum})  
gives also {\em diffuse flux} with normalization $E_s$ substituted by 
the energy density of the cascade radiation $\omega_{\rm cas}$.
Since the cascade spectra $n_\gamma(E_\gamma)$ are the same far 
all $E_s$ (apart from the total normalization) the spectra are the same
for any generation spectrum $Q_g(E_s)$. In other words the resultant
universal spectrum forgets about its parent generation spectrum.    

{\em The diffuse flux} of the cascade radiation is described by   
the space density of the cascade photons $n_\gamma(E_\gamma)$
with normalization given by Eq.~(\ref{eq:Knorm}) where $E_s$ is 
substituted by the energy density $\omega_{\rm cas}$.
The problem  here is an additional component at $E_\gamma >\calE_\gamma$ 
due to nearby sources (see subsecton \ref{sec:nearby}).

In conclusion, the  universality of spectrum (\ref{eq:gamma-spectrum}) 
for both remote point-like sources and diffuse radiation implies that 
all characteristics of the spectrum do not depend on the primary  
energy $E_s$ and distance $r$ (for a point-like source).
The universal spectrum, obtained above in a static universe, describes 
in fact rather wide class of remote gamma-ray sources in the universe 
and also diffuse gamma-ray radiation. In the next subsections~\ref{sec:comparison}, \ref{sec:large-z} and  \ref{sec:nearby} we will 
present the relevant comparison with numerical simulations for point-like 
sources and discuss the phenomena which limit the validity of the 
universal spectrum.

%%%%%%%%%%%%%%%%%%%%%%%%%%%%%%%%%%%%%%%%%%%%%%%%%%%%%%%%%%%%%%%%%%%%
\subsection{Comparison of universal spectrum with numerical simulations}
\label{sec:comparison}
In this subsection we compare the universal spectrum, obtained within 
our simplified model, with realistic numerical simulation.
Universal spectrum (\ref{eq:gamma-spectrum}) is not valid for sources at too
small and too large redshifts $z$. 
At small redshifts (nearby sources) absorption on EBL, which is
essential feature of our model, may not occur because of small 
distance and low space density of EBL photons. Cascades may not be
produced at all if a primary photon has energy $E_s$  less than
threshold  of pair production on EBL, $\calE_\gamma$. 

Note that in our calculations 
we did not include at all the density of target photons. Instead 
we just assumed that distance to a source is large enough for photon 
absorption at $E > \calE_\gamma$.

For nearby sources this assumption fails. In case of low 
density $n_{\rm ebl}$ of EBL photons, 
$\sigma_{\rm pair} n_{\rm ebl} r_{\rm source} \ll 1$, cascade may
still develop on CMB only; and this case will be considered below.   
Even if a photon is absorbed, cascade still may not develop. It 
happens when characteristic IC radiation length for secondary 
electron/ positron exceeds the distance to the observer. 

At the large redshits $z$  comparison with numerical simulations may
fail, because in analytic calculations we neglect effects of the 
universe expansion and energy redshift.  We will  include these effects  
in  the next subsection and thus extend our consideration to larger 
redshifts. 
\begin{figure}[h]\medskip
 \begin{minipage}[h]{\columnwidth}
 \includegraphics[width=\columnwidth]{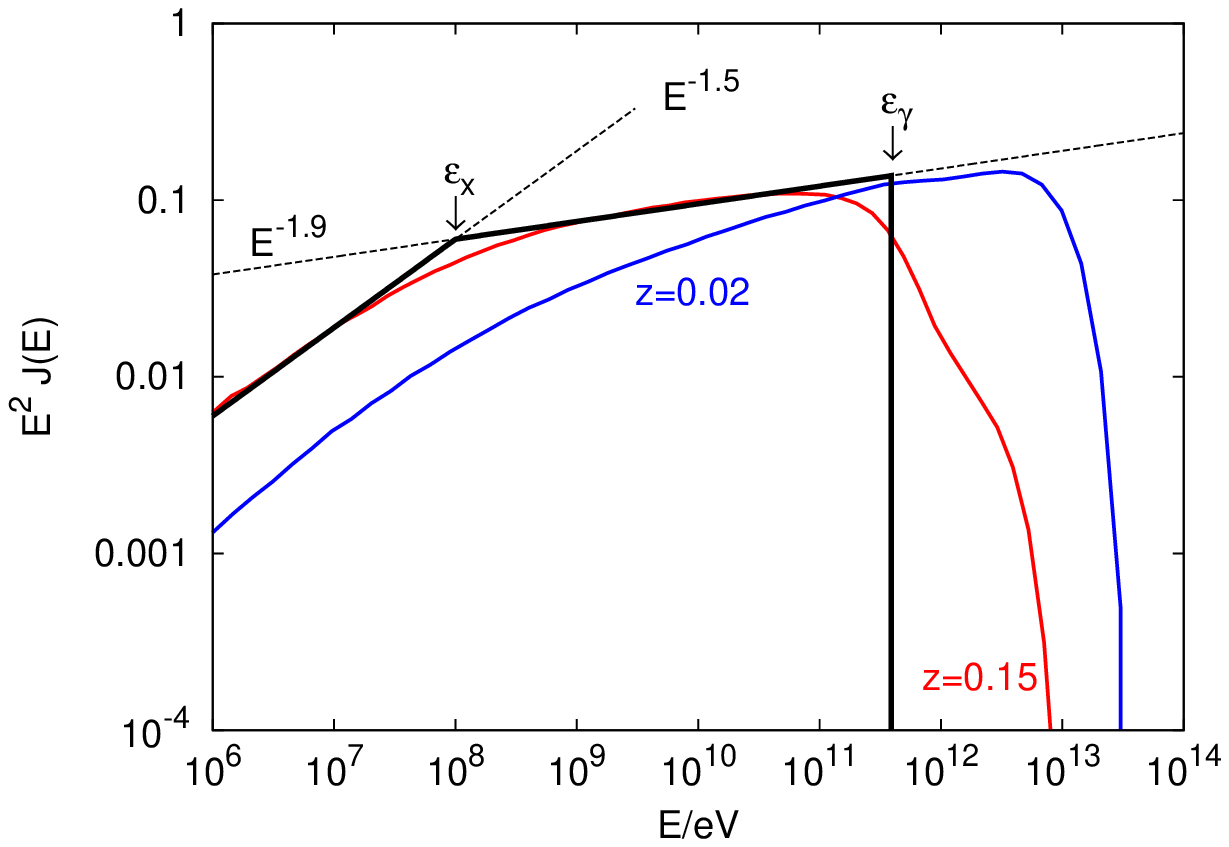}
 \end{minipage}
 \hspace{4mm}
 \begin{minipage}[h]{\columnwidth}
 \includegraphics[width=\columnwidth]{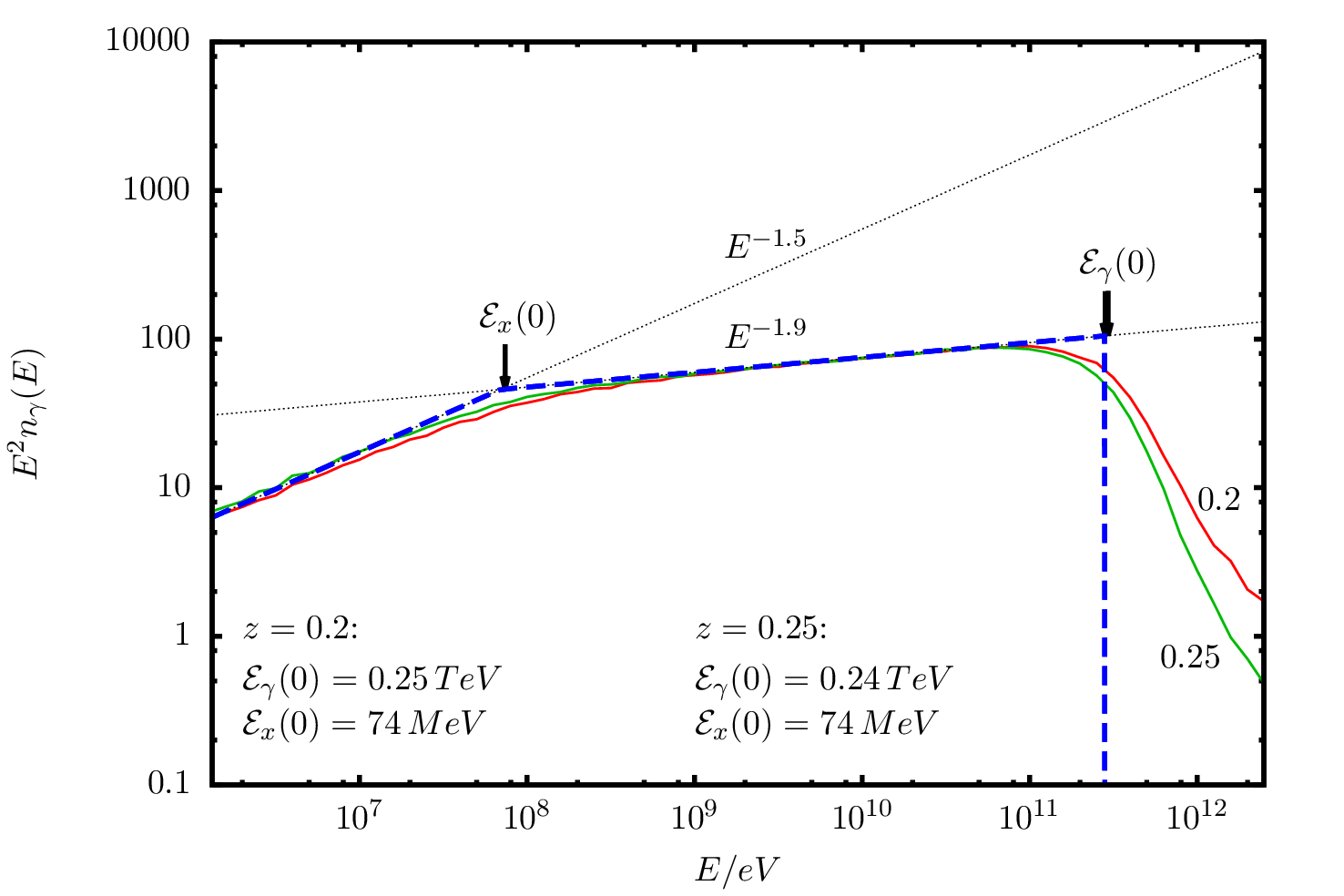} 
 \end{minipage}
 \caption{Comparison of analytic calculations (black thick line) 
with numerical simulation (red and blue lines for $z=0.15$ and $z=0.02$ 
correspondingly from the work by
Kachelrie\ss~et al~\cite{Kachelriess:2011bi} ({\em upper panel})
and from the present work ({\em lower panel}) for redshifts $z=0.2$ (red line)
and 0.25 (green line).  Analytic calculations in the lower panel
are presented by blue broken lines. 
For the both panels the line $\propto E^{-1.5}$ is shown for 
comparison; the characteristic energies in analytic calculations 
$\calE_X$ and $\calE_\gamma$ are shown by arrows (see the text).
} 
\label{fig:small_z}
\end{figure}
The first comparison of analytic calculations \cite{BGKO}    
(similar to our static-universe model) with MC simulation 
for sources with fixed redshift $z$ was done in \cite{Kachelriess:2011bi} 
and presented in the upper panel of Fig.~\ref{fig:small_z}.
We can compare now the predicted spectrum (\ref{eq:gamma-spectrum})  
shown by thick black line, with MC spectrum from \cite{Kachelriess:2011bi}.
For $z=0.15$~ ($r=626$~Mpc) we see the good agreement with the model where 
energy of EBL photons are fixed as $\eps_{\rm ebl}=0.676$~eV. The 
low-energy spectrum $\propto E^{-1.5}$ is reliably confirmed. 
The cascade multiplication spectrum ($\propto E^{-2}$) is seen as 
$E^{-1.9}$. Values of $\calE_X$ and $\calE_\gamma$ 
agree well with MC simulation. The MC simulation gives a smooth 
transition between different regimes, while in analytic calculations 
it is sharp because of dichromatic spectrum of background. 

For $z=0.02$~ $(r= 83.4~ Mpc)$ agreement is bad and it has to be
interpreted as small distance to the source.  

In the lower panel of Fig.~\ref{fig:small_z} the analytic spectrum
(\ref{eq:gamma-spectrum}) is compared with our numerical 
simulations with  the same conclusions (see subsection III). 

The success of static model is provided by large, but not too large
value of $z$. For larger redshifts we have to generalize our analytic
calculations for expanding universe. 
\begin{figure*}[h]
%\begin{center}
 \begin{minipage}[h]{60mm}
 \includegraphics[width=60mm,height=45mm]{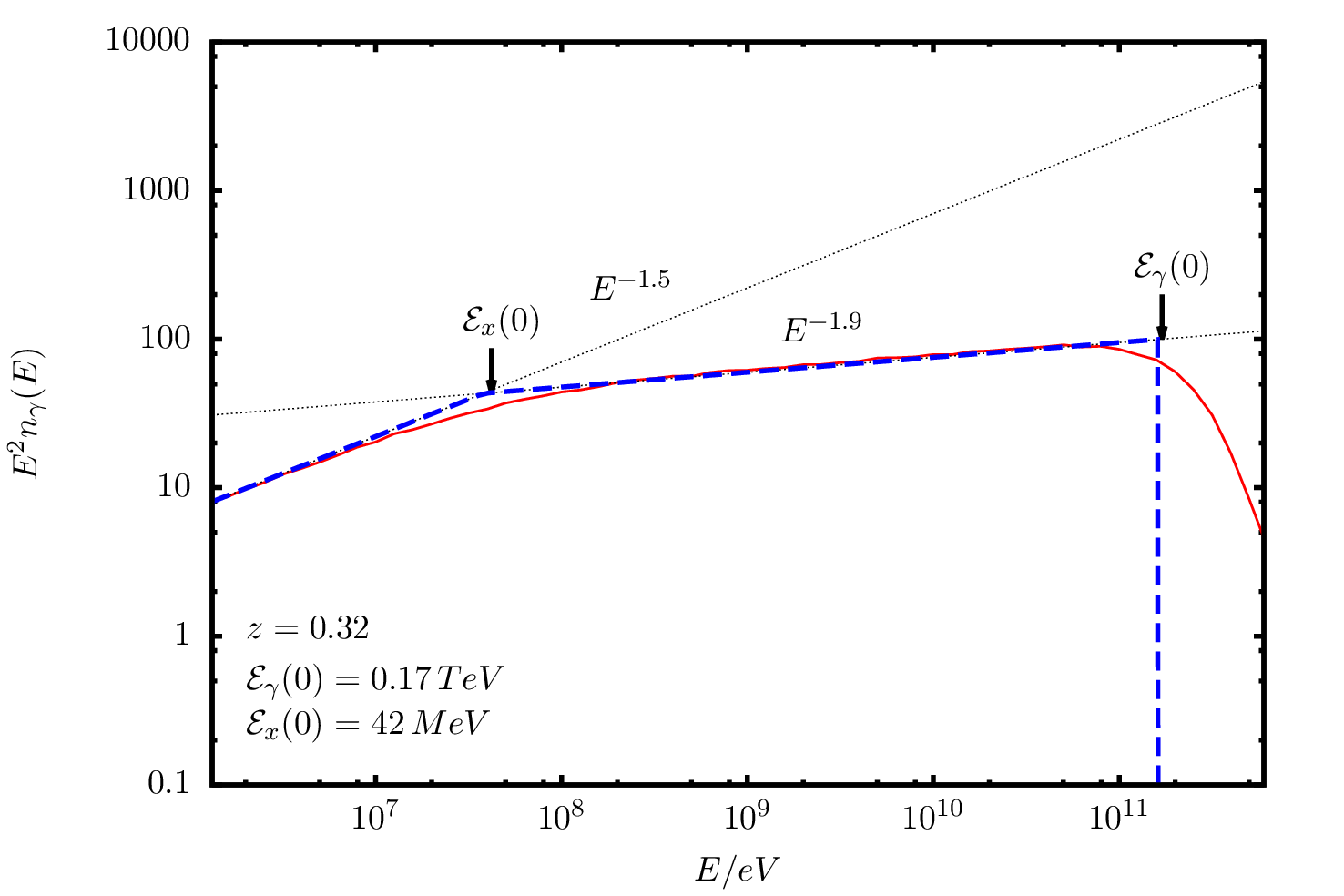}
 \end{minipage}
 \hspace{3mm}
 \begin{minipage}[h]{60mm}
 \includegraphics[width=60mm,height=45mm]{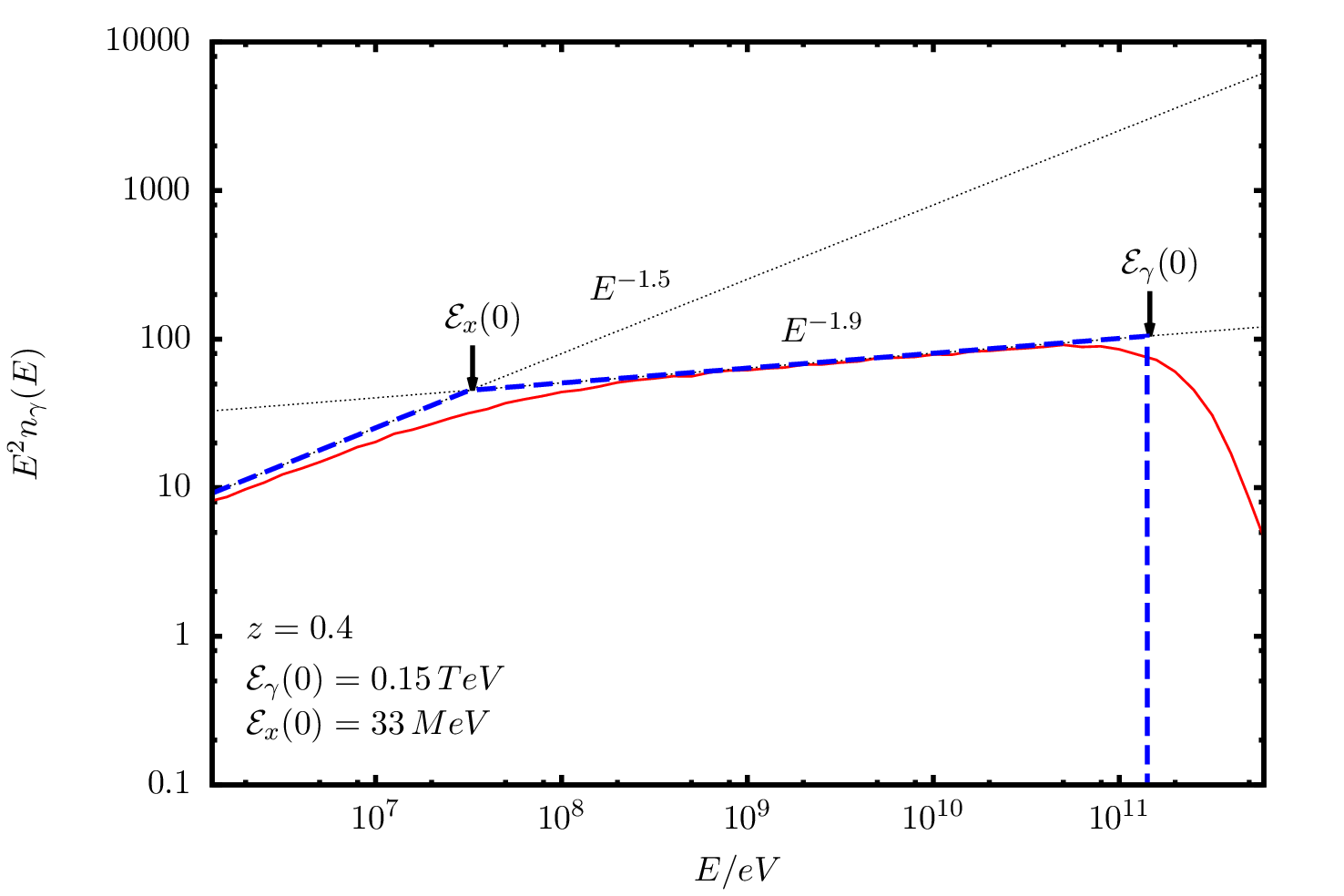}
 \end{minipage}
\newline \noindent
\medskip \hspace{-18mm}
 \begin{minipage}[h]{60mm}\vspace{2mm}
 \includegraphics[width=60mm,height=45mm]{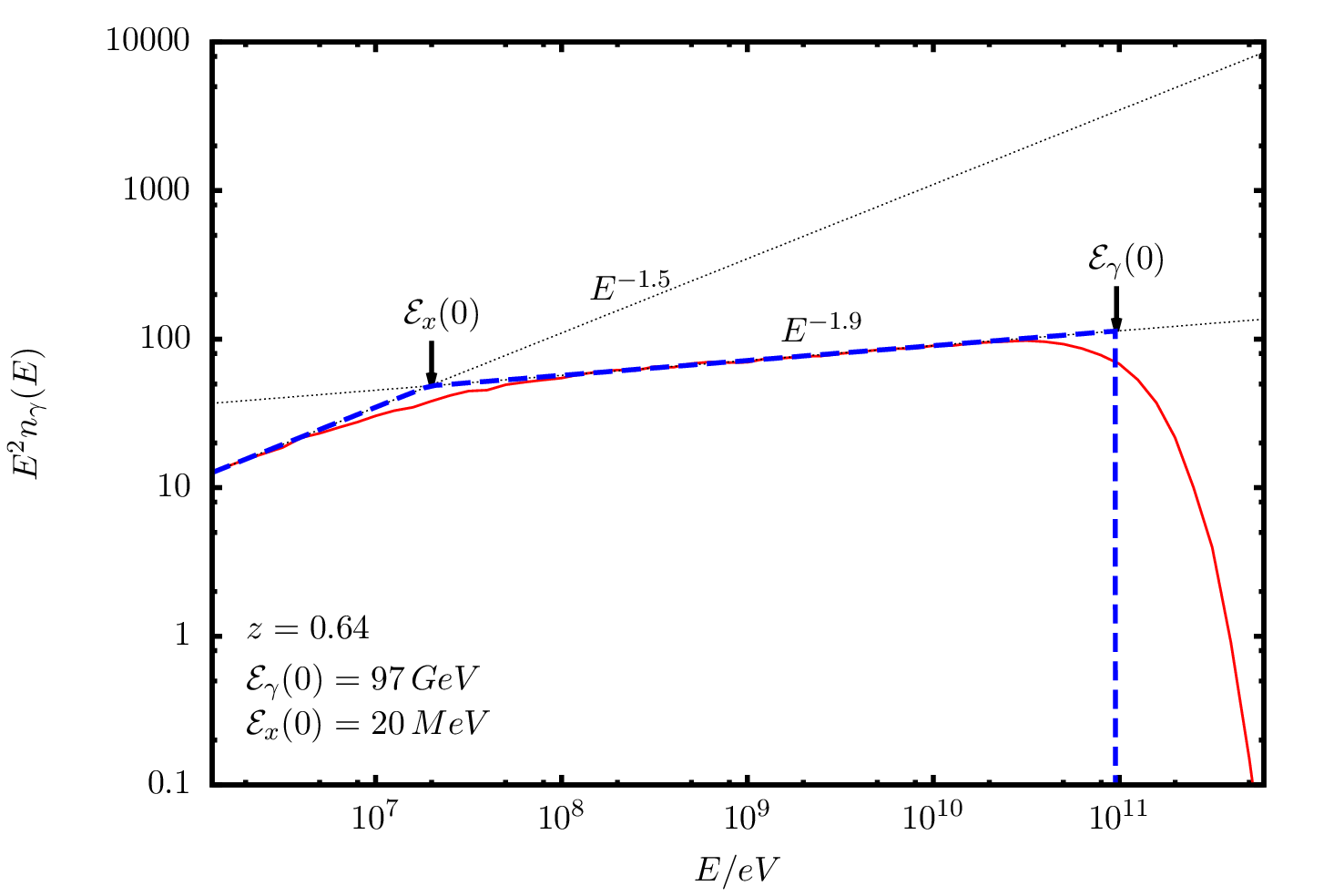}
 \end{minipage}
 \hspace{3mm}
 \begin{minipage}[h]{60mm}\medskip
 \includegraphics[width=60mm,height=45mm]{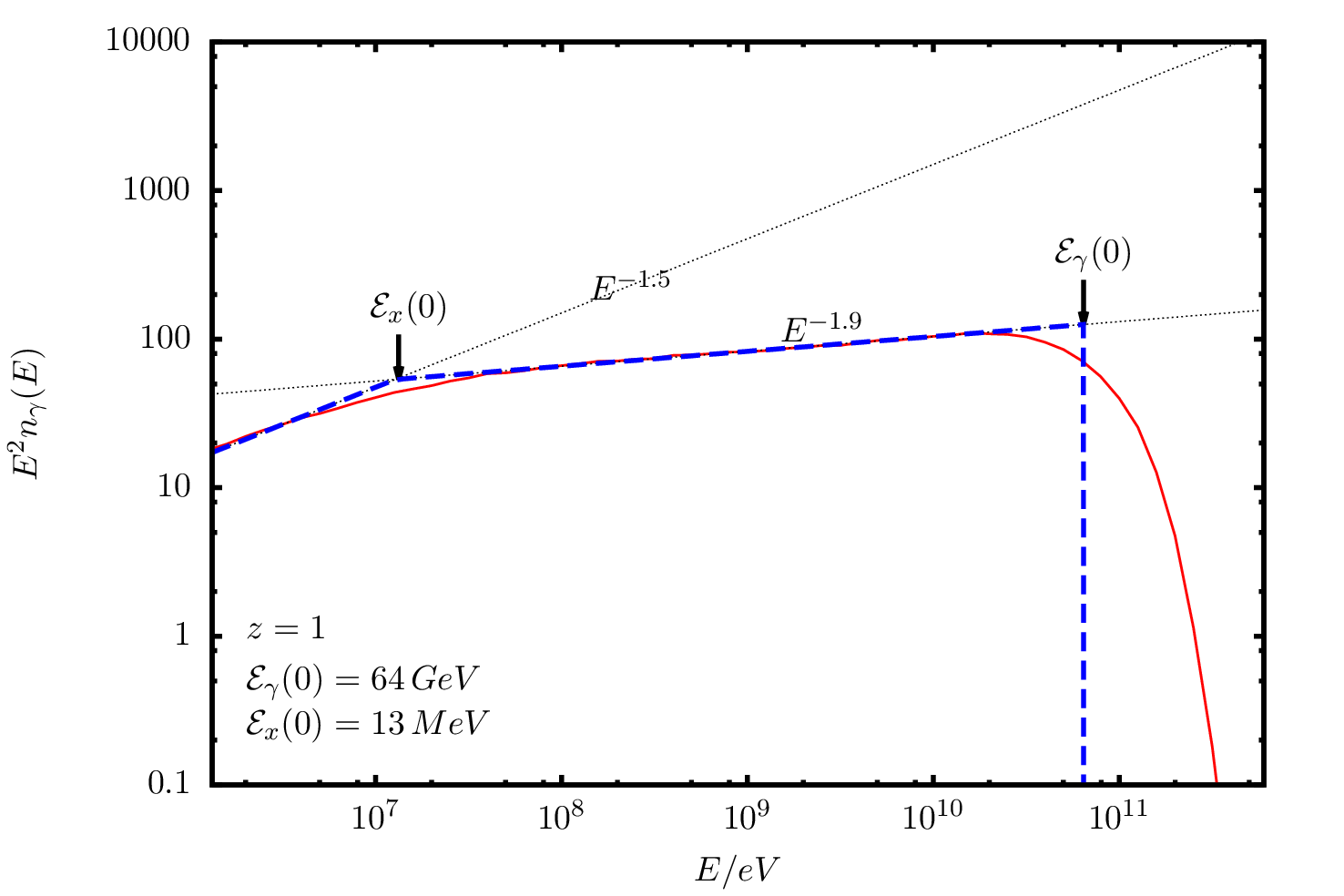}
\end{minipage}
 \vspace{-2 mm}%
\caption{ Comparison of analytic calculations with numerical
simulations from this work for large redshifts $z > 0.3$.  
Four panels show the comparison for redshifts 0.32,~ 0.4,~ 0.64,~ and 
1.0 with the same notation as in Fig.~\ref{fig:small_z}  (see the text 
for details).
} %
\label{fig:large_z}

\end{figure*} %
%

%%%%%%%%%%%%%%%%%%%%%%%%%%%%%%%%%%%%%%%%%%%%%%%%%%%%%%%%%%%%%%%%%%%
\subsection{Expanding universe and comparison for $z > 0.3$ }
\label{sec:large-z} 

For large redshifts $z > 0.3$ we assume that cascades have enough time 
to be developed during the Hubble time $H^{-1}(z)$ to the spectrum given by 
Eq.~(\ref{eq:gamma-spectrum}). The energies of background photons at epoch
$z$ are assumed to be $\eps_{\rm cmb}(z)=(1+z)\eps_{\rm cmb}$ for CMB
and $\eps_{\rm ebl}^z > \eps_{\rm ebl}$ for EBL. Accordingly, the
characteristic energies at redshift $z$  become 
$\calE_\gamma(z)=m_e^2/\eps_{\rm ebl}^z$ and 
$\calE_X(z)=\frac{1}{3}(\calE_\gamma(z)/m_e)^2\eps_{\rm cmb}(z)$. 
Spectrum of remnant photons at epoch $z$, $n_\gamma (E_\gamma,z)$, 
is given by Eq.~(\ref{eq:gamma-spectrum}) with 
$\calE_X \equiv \calE_X(z)$ and $\calE_\gamma \equiv \calE_\gamma(z)$. 
At propagation of this spectrum to $z=0$, the energies of 
photons $E_\gamma$ and characteristic energies $\calE_\gamma(z)$
and $\calE_X(z)$ are redshifted by factor $(1+z)$, but 
Eq.~(\ref{eq:gamma-spectrum}) remains the same with invariant value of 
$K$ provided by conservation of number of particles during the
redshift process, $n_\gamma(E_\gamma,z)dE_\gamma=n_\gamma(E,0)dE$. 
Redshifted characteristic energies are given by 
\be
\calE_\gamma(0)= \frac{m_e^2}{\eps_{\rm ebl}^z}\frac{1}{1+z}~,\;\;\;\:~
\calE_X(0)=\frac{m_e^2 \eps_{\rm cmb}}{3 (\eps_{\rm ebl}^z)^2} .
\label{eq:charact-ener}
\ee
From the second relation in Eq.~(\ref{eq:charact-ener}) one finds 
the energy of EBL photon $\eps_{\rm ebl}^z$ at epoch $z$ as 
\be
\eps_{\rm ebl}^z=\left (
\frac{m_e^2 \eps_{\rm cmb}}{3 \calE_X(0)} \right )^{1/2} .
\label{eq:ebl_z}
\ee
Now the procedure of comparing of analytic solution with numerical 
simulation consists of the the following four operations  illustrated
by Fig.~\ref{fig:large_z}: \\*[2mm] 
(i) Normalization of $E^{-1.5}$ part of analytic solution  by
$E^{-1.5}$ part of the simulation.\\
(ii) Finding $\calE_X(0)$ as intersection of $E^{-1.5}$ and $E^{-1.9}$ 
parts of the spectrum (see Fig.~\ref{fig:large_z}). In this
operation we neglect the difference  between $E^{-1.9}$ and $E^{-2}$.\\
(iii) Calculation of $\eps_{\rm ebl}^z$ using Eq.~(\ref{eq:ebl_z}).\\
(iv) Calculation of cutoff energy $\calE_\gamma (0)$ from 
Eq.~(\ref{eq:charact-ener}).

Comparison of analytic solutions with numerical simulation is
presented in Fig.~\ref{fig:large_z} for redshifts $z = 0.32,~ 0.4,~
0.64,~ 1.0$. The analytic solution predicts that cascade spectrum 
consists of two power-law components $E^{-1.5}$ and $E^{-2}$, with
intersection
at $\calE_X$.  The numerical simulations confirm this prediction with 
component $E^{-1.9}$ instead of $E^{-2}$. The only free parameter in 
analytic calculations is energy of EBL photons for each redshift,
which means the value of $\eps_{\rm ebl} \approx 0.68$ at $z=0$ and 
evolution $\eps_{\rm ebl}^z/\eps_{\rm ebl}=f(z)$. Free parameter 
$\eps_{\rm ebl}^z$ determines two characteristic energies   
$\calE_X$ and $\calE_\gamma$. 

In contrast to the sharp spectral features in analytic calculations ,
the numerical simulations predict a smooth transition
regimes centered by these features. 

The evolution of EBL energy $\eps_{\rm ebl}^z/\eps_{\rm ebl}$ 
with $z$ in dichromatic model must correspond, to some extent, to the
evolution of the mean EBL energy $\bar{\eps}_{\rm ebl}(z)$ with redshift.

\begin{figure}[htb]
\centering\includegraphics[angle=0,width=0.5\textwidth]{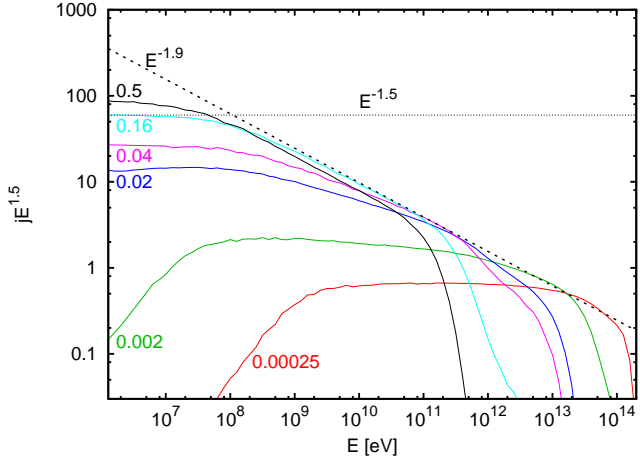}
\caption{ Merging at increasing $z$ of numerically-simulated
spectra to the universal $E^{-1.9}$ spectrum, which within accuracy 
of calculations coincides with fundamental cascade-multiplication 
spectrum $\propto E^{-2}$ in analytic calculations. 
}
\label{fig:merging}
\end{figure}

The numerical simulations show an interesting feature of merging the 
spectrum $n_\gamma(E,z)$ to $E^{-1.9}$ at increasing $z$, which 
coincides within accuracy of calculations with fundamental spectrum of 
the cascade multiplication $E^{-2}$ in analytic calculations 
(see Fig.~\ref{fig:merging}).   
%

%%%%%%%%%%%%%%%%%%%%%%%%%%%%%%%%%%%%%%%%%%%%%%%%%%%%%%%%%%%%%%%%%%
\subsection{Cascading only on CMB}
\label{sec:cascadeCMB}
This section has a technical character. Study of the cascade development  
only  on CMB is important at least in two cases discussed in this paper:

(i) In the case of nearby sources when absorption of photons on EBL is absent.

(ii) In expanding universe at large redshifts, when due to high space 
density of CMB photons a cascade is developing very fast with cutoff at 
$\calE_\gamma^{\rm cmb}= m_e^2/\eps_{\rm cmb}^z$.  Only later the 
HE tail of the cascade photon distribution is slowly absorbed on 
EBL, followed by IC of produced electrons and positrons on CMB photons. 
  
Consider first cascading on {\em monochromatic CMB} at $z=0$. The
energy spectrum of cascade 
is given again by Eq.~(\ref{eq:gamma-spectrum}) with 
\begin{equation}
\calE_\gamma = \calE_\gamma^{\rm cmb} =m_e^2/\eps_{\rm cmb}
\approx 4.1\times 10^{14}~{\rm eV} 
\label{eq:E_gamma}
\end{equation}
and
\begin{equation}
\calE_X = (4/3)(\calE_e/m_e)^2\eps_{\rm cmb}
=(1/3)\calE_\gamma^{\rm cmb}
%%=1.38 \times 10^{14}~{\rm eV}
\label{eq:E_X}
\end{equation}
The normalization is given by the same equation (\ref{eq:Knorm}) 
with $\calE_\gamma/\calE_X=3$ as follows from Eq.~(\ref{eq:E_X}).

The remarkable feature of this calculation is prediction of very
narrow cascade-multiplication energy width given by ratio 
$\calE_\gamma/\calE_X = 3$, to be compared with $3\times 10^3$ 
for universal spectrum (\ref{eq:gamma-spectrum}). This is the direct 
consequence of monochromatic CMB model accepted here (see
discussion in the end of this subsection).

For redshift $z$ there are two spectra of interest: the equilibrium
spectrum at epoch $z$, $n_\gamma(E_\gamma,z)$, and this spectrum 
redshifted to epoch $z=0$, $n_\gamma(E,0)$ . 

Consider first the former. It is given by Eq.~(\ref{eq:gamma-spectrum}) 
with $\eps_{\rm cmb}^z=\eps_{\rm cmb}(1+z)$ and by characteristic
energies at epoch $z$ as
\be
\calE_\gamma(z)=m_e^2/\eps_{\rm cmb}^z,\;\;\;\;
\calE_X(z)= (1/3)\calE_\gamma(z) .
\label{eq:char-z}
\ee
The normalization is given by Eq.~(\ref{eq:Knorm}) with 
$\calE_\gamma(z)/\calE_X(z)=3$. 

The spectrum redshifted to $z=0$ may be also of interest for
applications. It is given by redshifted characteristic energies 
$\calE_\gamma(0)=\calE_\gamma(z)/(1+z)$ and 
$\calE_X(0)=\calE_X(z)/(1+z)$:
\be
\calE_\gamma(0)=
\frac{m_e^2}{\eps_{\rm cmb}(1+z)^2}
\label{eq:Egamma0}
\ee
and 
\be
\calE_X(0)=(1/3)\calE_\gamma(0)=(1/3)\frac{m_e^2}{\eps_{\rm cmb}(1+z)^2}
\label{eq:EX(0)}
\ee
The redshift leaves constant $K$ in normalization (\ref{eq:Knorm})
invariant.

In fact, the monochromatic CMB model considered above is unrealistic and 
introduction of dichromatic model is necessary in most applications. 
It can be explained for example by cascading on CMB at $z=0$ at 
distance $r$ from an observer. IC scattering occurs on all CMB photons
which energy can be assumed as $\eps_{\rm cmb}=6.3\times 10^{-4}$~eV. 
However, absorption of photons occurs on CMB photons from high energy 
tail of Planckian distribution and minimum energy of absorbed photons 
$\calE_{\min}$ is determined by much higher energies of CMB photons 
$\tilde{\eps}_{\rm cmb} \gg \eps_{\rm cmb}$ as 
\be
\tilde{\calE}_{\min}=\tilde{\calE}_{\min}^{\rm cmb}=m_e^2/\tilde{\eps}_{\rm cmb}.
\label{eq:Emin_cmb}
\ee
Energy $\tilde{\eps}_{\rm cmb}$ coincides with minimum energy in
HE tail of CMB photons for given $E_{\gamma}$ because with increasing 
of $\eps$ the number density of CMB photons exponentialy falls down. 

Therefore, we come back to the standard spectrum given by 
Eq~(\ref{eq:gamma-spectrum}) distorted at the highest energies due to
distortion of relation $E_\gamma=\frac{4}{3}\gamma_e^2\eps_{\rm cmb}$
and with $\eps_{\rm ebl}$ substituted everywhere by $\tilde{\eps}_{\rm cmb}$.
\begin{figure*}[h]
\begin{center}
 \begin{minipage}[h]{60mm}
 \includegraphics[width=60mm,height=45mm]{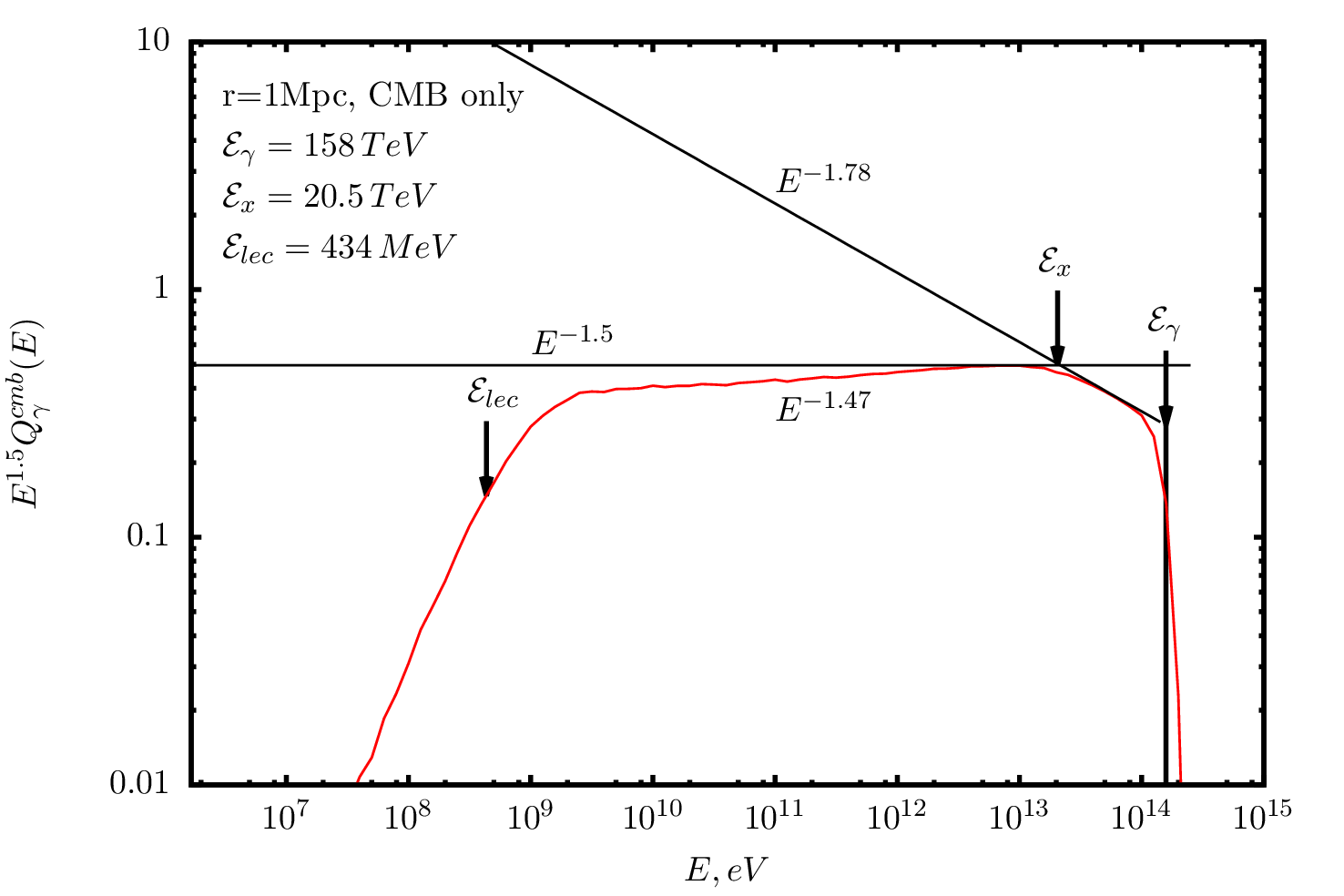}
 \end{minipage}
 \hspace{3mm}
 \begin{minipage}[h]{60mm}
 \includegraphics[width=60mm,height=45mm]{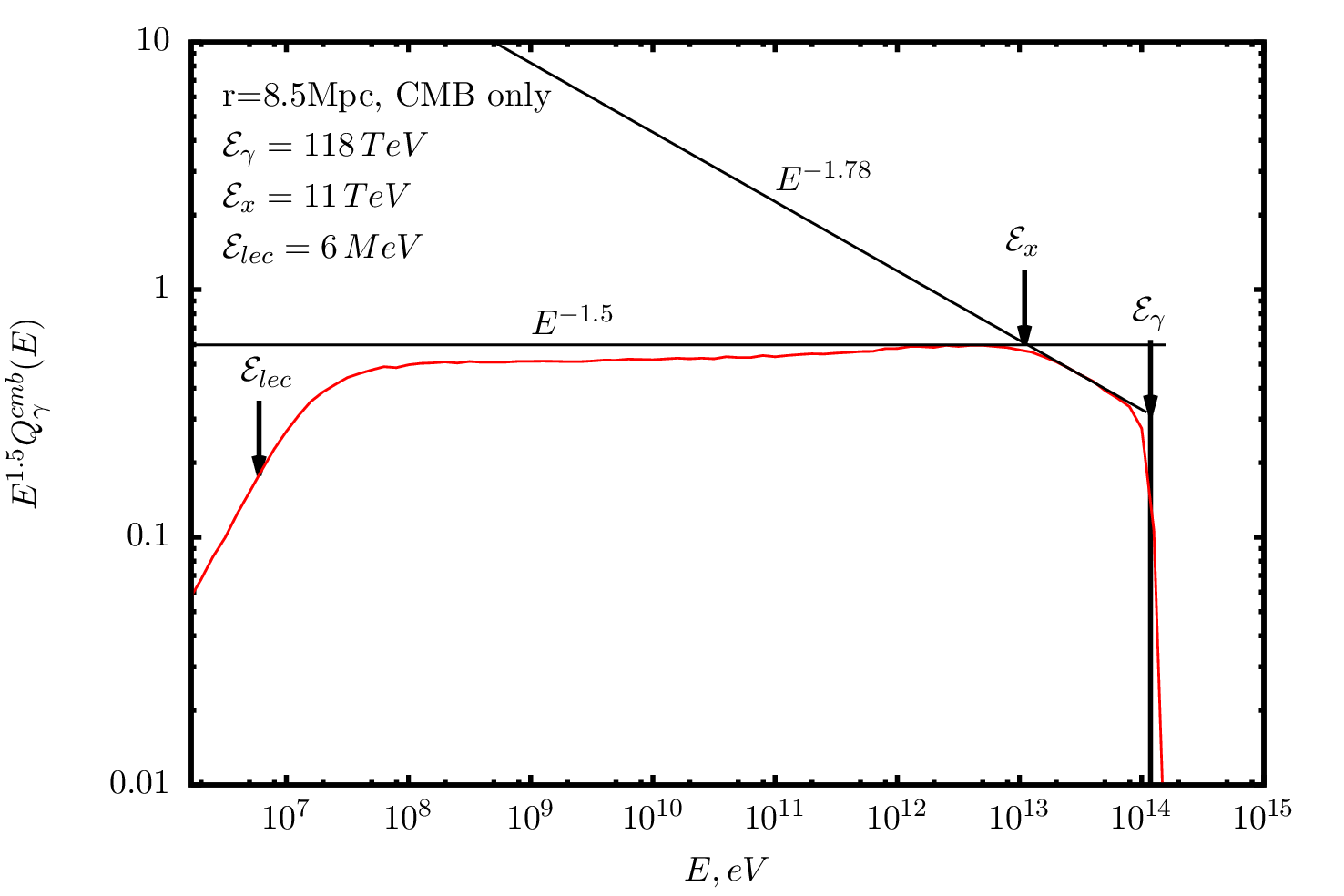}
 \end{minipage}
\newline \noindent
\medskip \hspace{-19mm}
 \begin{minipage}[ht]{60mm}\vspace{2mm}
 \includegraphics[width=60mm,height=45mm]{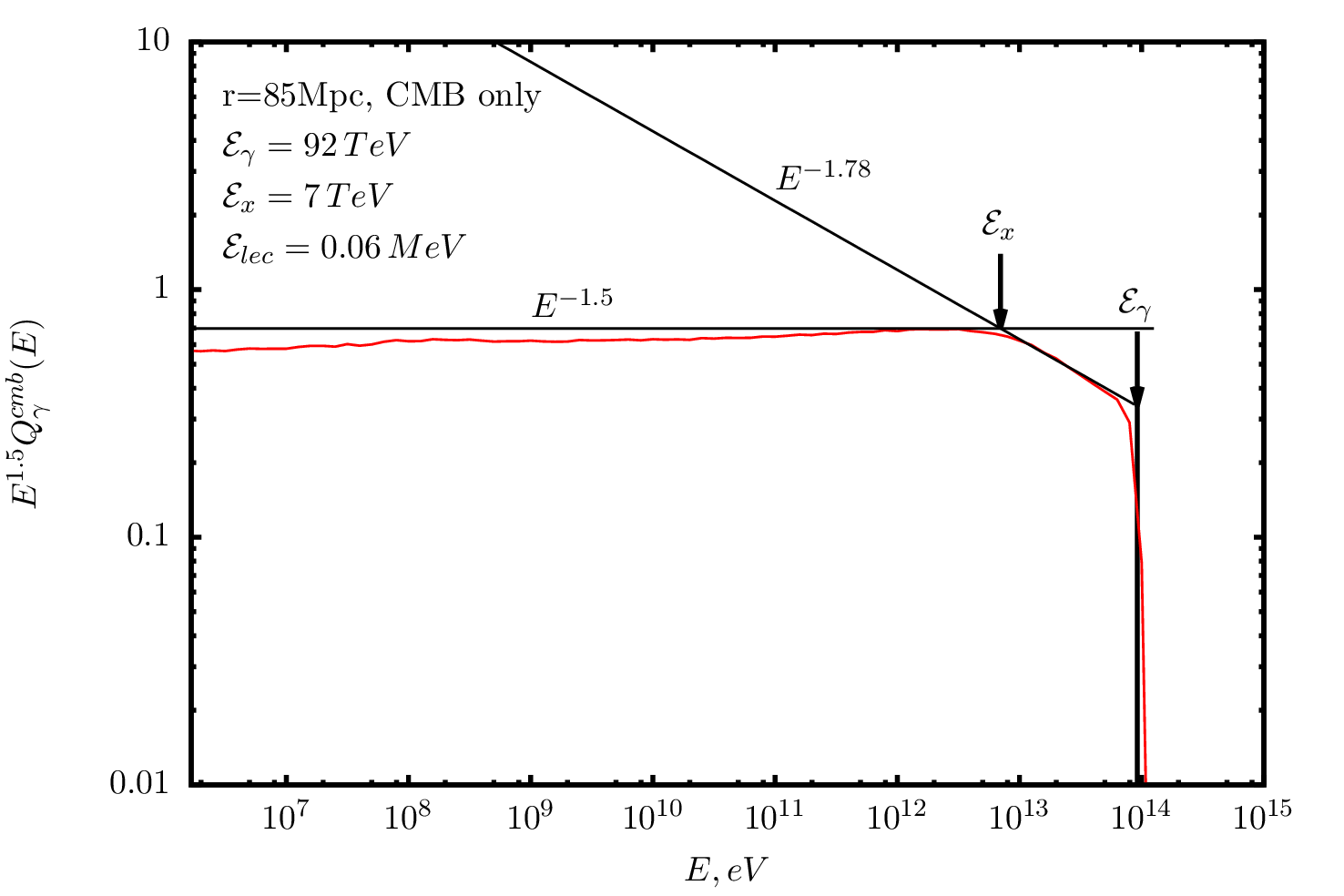}
 \end{minipage}
 \hspace{3mm}
 \begin{minipage}[h]{60mm}\medskip
 \includegraphics[width=59mm,height=45mm]{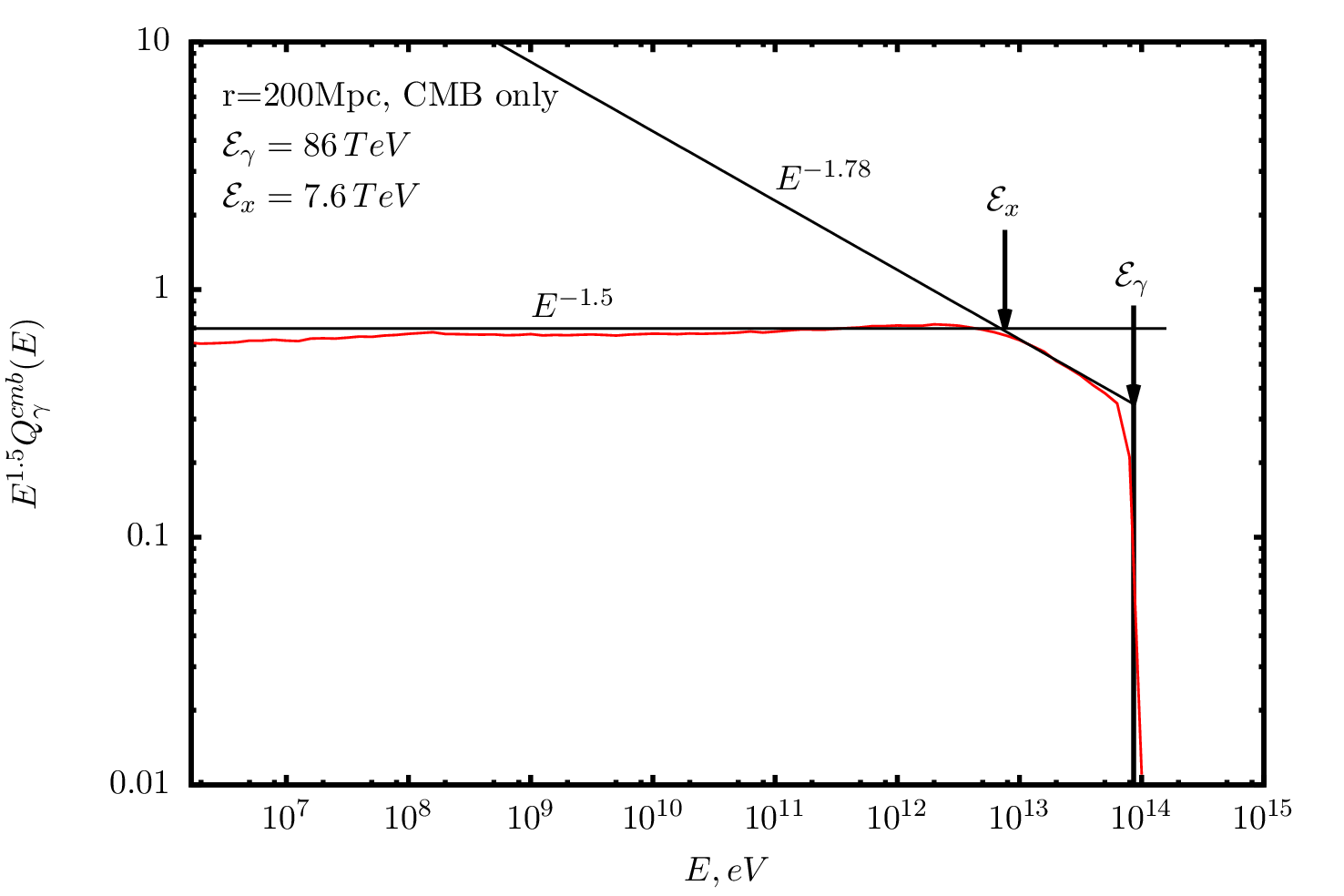}
\end{minipage}
 \vspace{-2 mm}%
\caption{ Comparison of analytic calculations with numerical
simulations for CMB radiation only for nearby sources at distances 
1.0~Mpc,~ 8.5~Mpc, 85~Mpc and 200~Mpc. 
} %
\label{fig:nearby_cmb}
\end{center}
\end{figure*} %
%
%%%%%%%%%%%%%%%%%%%%%%%%%%%%%%%%%%%%%%%%%%%%%%%%%%%%%%%%%%%%%%%%%%%%
\subsection{Cascades from nearby sources}
\label{sec:nearby} 
In this section we calculate the cascade spectrum taking into
account only CMB radiation and compare them with  numerical calculations 
at the same assumptions. This procedure clarifies the  interesting 
physical details.
As  argued above the cascade spectrum on CMB-only must be
calculated using the dichromatic model of background radiation 
with $\eps_{\rm cmb}=6.3\times 10^{-4}$~eV, which provides IC
scattering, and higher energy $\tilde{\eps}_{\rm cmb}$, which
provides absorption of cascade photons with minimum energy
$\tilde{\calE}_{\min}$ given by Eq.~(\ref{eq:Emin_cmb}). 
As physics is concerned $\tilde{\eps}_{\rm cmb}$ is minimal 
energy in HE Planckian tail of photon distribution where number 
density of photons is big enough enough for absorption of photons  
with energy $\tilde{\calE}_{\min}$. The energy $\eps_{\min}$ for  
$\tilde{\calE}_{\min}$ can be calculated from kinematic relation 
for arbitrary $E_\gamma$ and $\eps$ given by 
$E_{\gamma}\eps (1+\cos\phi)=2\varepsilon_{cm}^2$, where $\phi$
is an angle between photons $E_\gamma$ and $\eps$ in laboratory 
system and $\varepsilon_{cm}$ is energy of each photon in cm-system. 
Finding the value $\eps$ from this relation for 
$E_\gamma= \tilde{\calE}_{\min}$ and minimizing it with the choice 
$\cos\phi=1$ and $\varepsilon_{cm}= m_e$, we arrive at 
$\eps_{\min}=m_e^2/\tilde{\calE}_{\min}$ as in Eq.~(\ref{eq:Emin_cmb}).

We can discuss now the spectra and basic features predicted for 
the cascades on CMB from nearby sources, assuming tentatively, as the
first step, $\eps_{\rm cmb} \sim 6\times 10^{-4}$~eV and 
$\tilde{\eps}_{\rm cmb} \sim 2\times 10^{-3}$~eV (these values will be 
determined more precisely after comparison with numerical simulations.) 

The minimum absorption energy (on CMB photons) can be estimated 
from equation $\ell_{\rm abs}^{\rm cmb}(\tilde{\calE}_{\min})=r$, where 
$\ell_{\rm abs}^{\rm cmb}(E_\gamma)$ is absorption  length in CMB 
background. It results in well known value 
$\tilde{\calE}_{\min} \sim 1\times 10^{14}$~eV and it is given precisely,  
as $\tilde{\calE}_\gamma$, for different distances $r$ in Table \ref{table1}. 
Using Eq.~(\ref{eq:Emin_cmb}) we find then 
$\tilde{\eps}_{\rm cmb}\sim 2\times 10^{-3}$~eV. The transition energy
is given by $\calE_X=(1/3)(\tilde{\calE}_{\min}/m_e)^2\eps_{\rm cmb}$ which
approximately equals to $7\times 10^{12}$~eV. We cannot expect in this 
picture the standard spectra $\propto E^{-1.5}$ and $\propto E^{-2}$,
especially the latter one. 
First of all the spectrum can be flatter than $\propto E^{-2}$  
due to flattening of $E_\gamma \propto E_e^{-2}$ regime in the end of
the IC spectrum (see Fig.~\ref{fig:EsecGamma}).
 
Another plausible reason is 
the width of the cascade-multiplication part of the spectrum, i.e. 
one with conventional $E^{-2}$ spectrum, but with ratio 
$\calE_\gamma/\calE_X \approx 14$ to be compared with $3\times 10^3$ 
for the case of $(\eps_{\rm cmb},~\eps_{\rm ebl})$ background.  
The relatively small ratio $\calE_\gamma/\calE_X \approx 14$ is the
reminiscent of the ratio 3 for CMB monochromatic background model 
(see Eq.~(\ref{eq:E_X})). Since the realistic simulations show a 
smooth transition between energies $\calE_X$ and $\calE_\gamma$, one may
expect non-power-law of cascade-multiplication spectrum, in disagreement 
with the dichromatic model. As for low-energy asymptotic  
spectrum $E^{-1.5}$, it also should be distorted, because absorption 
on $\tilde{\eps}_{\rm cmb}$ photons occurs at distances of order of 
distance to the source $r$ and thus a cascade is underdeveloped. 
Now we will compare the analytic estimates with
accurate numerical simulations for cascades on CMB radiation,
and obtain more precisely parameters of our dichromatic model 
for different distances $r$ to the source. We will change the 
notation as $\tilde{\calE}_{\min} \equiv \tilde{\calE}_\gamma$ 
to emphasize that this is the value of spectrum cutoff. 

As the first step we find $\tilde{\calE}_\gamma$ from
equation $\ell_{\rm abs}^{\rm cmb}(\tilde{\calE}_{\gamma})=r$, where 
$\ell_{\rm abs}^{\rm cmb}(E_\gamma)$ is absorption  length in CMB 
background. The values of $\tilde{\calE}_{\gamma}$ are shown in Table 
\ref{table1} for distances 1.0~, 8.5~, 85~, and 200~ Mpc. 

Next we calculate $\tilde{\eps}_{\rm cmb}$ using Eq.~(\ref{eq:Emin_cmb})
and values of $\tilde{\calE}_{\min} \equiv \tilde{\calE}_{\gamma}$
from the Table~\ref{table1} (third row).

The values of $\calE_X$ in the fourth row are obtained from comparison with 
numerical simulations (see Fig.~\ref{fig:nearby_cmb} ) as intersection 
of the power-law approximation of the cascade-multiplication spectrum 
($\propto E^{-1.8}$ in the Fig.~\ref{fig:nearby_cmb}) with low-energy
asymptotic. 

The energies of CMB photons $\eps_{\rm cmb}$ (fifth row) are calculated
using the equation $\calE_X=(1/3)(\calE_\gamma/m_e)^2\eps_{\rm cmb}$.

In the last row of Table~\ref{table1} we put the low-energy 
cutoff $\calE_{\rm lec}^{\gamma}$ of the cascade spectrum  
estimated in the following way. 

The low-energy cascade electrons with energy below some critical energy 
$E_e^{\rm cr}(r)$ have a time of IC photon emission larger than 
time-of-flight $r/c$. Therefore radiation of IC photons with 
energies below 
$\calE_{\rm lec}^{\gamma}=(4/3)(E_e^{\rm cr}/m_e)^2\eps_{\rm cmb}$ 
is suppressed. 

The critical energy of electron $E_e^{\rm cr}$ can be found from 
$\tau_e^{-1}(E_e^{\rm cr})=c/r$, where $\tau_e(E)$ is the electron lifetime 
relative to IC energy loss: 
\be
\tau_e^{-1}(E_e)=\left (\frac{1}{E_e}\frac{dE_e}{dt}\right )_{\rm IC}=
\frac{4}{3}\sigma_Tc\gamma_e\frac{\rho_{\rm cmb}}{m_e},
\label{eq:ICloss}
\ee
where $\sigma_T$ is the Thompson cross-section, $\gamma_e=E_e/m_e$ is the 
electron Lorentz-factor and $\rho_{\rm cmb}=\eps_{\rm cmb}n_{\rm cmb}$
is energy density of CMB radiation. Eq.~(\ref{eq:ICloss}) can be
rearranged as 
\be
\tau_e^{-1}(E_e)=
\frac{4}{3}\sigma_T c n_{\rm cmb}\frac{E_e}{\calE_\gamma^{\rm cmb}}
\label{eq:tau_e}
\ee
where $\calE_\gamma^{\rm cmb}=m_e^2/\eps_{\rm cmb}$.

Using $\tau_e^{-1}(E_e^{\rm cr})=c/r$ one finds the critical energy of 
electron $E_e^{\rm cr}$ and low-energy cutoff  $\calE_{\rm lec}^\gamma$ as 
\be
E_e^{\rm cr}=\frac{3/4}{\sigma_T n_{\rm cmb} r}\calE_\gamma^{\rm cmb};~~
\calE_{\rm lec}^\gamma=\frac{3}{4}\left (\frac{1}{\sigma_T n_{\rm cmb}r}
\right )^2\calE_\gamma^{\rm cmb}
\label{eq:Elec}
\ee

From Table \ref{table1} one can see that for nearby sources at
distance 1 - 85 Mpc the dichromatic model is characterized  by  
almost equal energies $\eps_{\rm cmb} \approx 6\times 10^{-4}$~ eV
and by similar values of 
$\tilde{\eps}_{\rm cmb}\approx (2 - 3)\times 10^{-3}$~eV.  For 200~Mpc 
these values differ more considerably. The low-energy cutoff is observable 
only for very close sources $\calE_{\rm lec}^\gamma \sim 400$~MeV 
for $r=1$~Mpc; for $r=8.5$~Mpc it starts at 6~MeV.

 Finally we calculate the cascade-photon spectrum taking the characteristic
energy features from Table~\ref{table1} and compare this spectrum
with precise numerical simulation on CMB radiation only. One may
expect that canonical low-energy part of spectrum $\propto E^{-1.5}$ 
will survive for long-distance sources and may fail for short-distance
ones, being underdeveloped. The Fig.~\ref{fig:nearby_cmb} confirms
this expectation: for distance $r=200$~Mpc the spectrum coincides well 
with $E^{-1.5}$ shape, and weakly distorted at smaller distances.  
For high-energy part of the spectrum $(\calE_X - \calE_\gamma)$ the
energy interval is very short and spectrum is $\propto E^{-1.78}$ 
flatter than canonical $2.0$ .

%%%%%%%%%%%%%%%%%%   Table         %%%%%%%%%%%%%%%%%%%%%%%%%%%%%%%%%%%%%%%
\begin{table}[t!]
\begin{center}
\caption{Parameters of analytic model for nearby sources (CMB only).}
\begin{tabular}{c|c|c|c|c|c}
\hline
distance $r$  &1~Mpc     & 8.5~ Mpc & 85~Mpc & 200~Mpc \\
\hline
$\tilde{\calE}_\gamma$~ eV & $1.58\times 10^{14}$ & $1.18\times 10^{14}$ &
$9.24\times 10^{13}$ & $8.6\times 10^{13}$ \\
\hline
$\tilde{\eps}_{\rm cmb}$~ eV & $1.65\times 10^{-3}$ & $2.21\times 10^{-3}$
& $2.83\times 10^{-3}$ & $3.04\times 10^{-3}$ \\
\hline
$\calE_X$~ eV & $2.05 \times 10^{13}$ & $1.11 \times 10^{13}$
& $7.0\times 10^{12}$ & $7.6\times 10^{12}$\\
\hline
$\eps_{\rm cmb}$~ eV & $6.42\times 10^{-4}$ & $6.24 \times 10^{-4}$ 
& $6.43 \times 10^{-4}$ & $8.0\times 10^{-4}$\\
\hline
$\calE_{\rm lec}^\gamma$~ eV & $4.34 \times 10^8$ & $6.01 \times 10^6$
& $6.01\times 10^4$ & $1.09 \times 10^4$\\
\hline
\end{tabular}
\label{table1}
\end{center}
\end{table}
%%%%%%%%%%%%%%%%%%%%%%%%%%%%%%%%%%%%%%%%%%%%%%%%%%%%%%%%%%%%%%%%%%%%%%

\section{Numerical simulations of electromagnetic cascade propagation}
\label{sec:simul}
In this section we will discuss the basics of numerical simulations,
universality of em-cascade spectra in the numerical simulations, the
calculated cascade energy spectra and upper limits on cascade energy 
density $\omega_{\rm cas}$ obtained from comparison of the calculated 
spectra with observations of Fermi LAT.

%%%%%%%%%%%%%%%%%%%%%%%%%%%%%%%%%%%%%%%%%%%%%%%%%%%%%%%%%%%%%%%%%%%%%
\subsection{Generalities}
\label{sec:generalities}

\begin{figure}[t]
\begin{center}
\includegraphics[angle=0,width=0.5\textwidth]{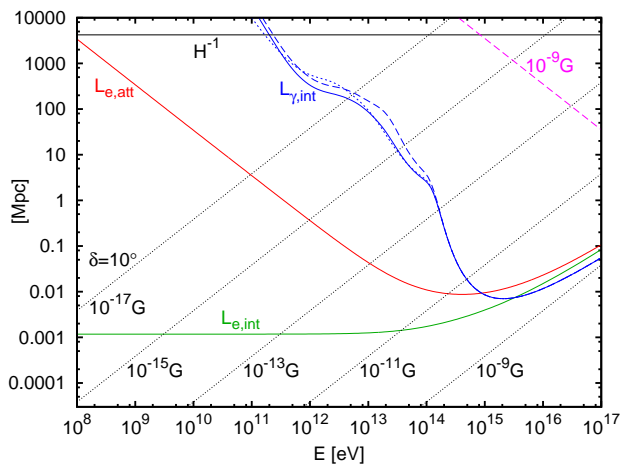}
\end{center}
\caption{ Shown are: pair production interaction length (see 
Eq.~\ref{int_length}) 
calculated assuming EBL models of Ref.~\cite{Kneiske:2003tx} (shown by 
solid blue line), of Ref.~\cite{Kneiske:2010pt} (dashed blue line) and 
of Ref.~\cite{Inoue:2012bk} (doted blue line); electron attenuation length 
(see Eq.~\ref{att_length}) due to inverse Compton scattering (red 
line) and interaction length of this process (green line); $10^{\circ}$ 
deflection length for electrons (shown by black dotted lines) for given 
values of constant transverse component of magnetic field; synchrotron 
energy loss length (see Eq.~\ref{synchrotronAttLength}) of electrons in 
$B_{\perp}=10^{-9}$G (shown by pink dashed line); adiabatic energy loss 
length (shown by horizontal solid black line). } 
\label{rates}
\end{figure}
%%%%%%%%%%%%%%%%%%%%%%%%%%%%%%%%%%%%%%%%%%%%%%%%%%%%%%%%%%%%%%%%%%%
The results presented in this work have been obtained with two independent 
numerical techniques, the Monte Carlo simulation and the 
code~\cite{okPHD,Gelmini:2011kg} based on solution of Boltzmann kinetic 
equations for cascade particles propagation in one dimension. The latter 
method doesn't take into account deflections of cascade particles by 
magnetic field and therefore is valid only for calculations 
 with averaged angles and time or for diffuse fluxes.
In Fig.~\ref{rates} interaction and energy loss lengths are shown for 
electrons and photons. The lengths are defined as follows:
\begin{eqnarray}
&& L^{-1}_{\rm i,int} = \int d\epsilon\,n(\epsilon) \int d\mu\frac{1-\beta_i 
\mu}{2} \sigma_i \label{int_length} \\
&& L_{\rm e,att} = L_{\rm e,int} E_{\rm e}/\bar{E}_{\gamma}, 
\label{att_length}
\end{eqnarray}
where $i=e,\gamma$; $\sigma_e=\sigma_{IC}$; 
$\sigma_{\gamma}=\sigma_{\rm PP}$ and $\bar{E}_{\gamma}$ is the mean 
energy of the recoil photon in IC.
A number of kinetic equation-based codes has been developed at present 
(see e.g.~\cite{Lee:1996fp,Yoshida:1998it,okPHD}).
For precise calculation of $\gamma$-ray fluxes from individual sources 
in presence of non-negligible magnetic fields, full 3D Monte Carlo 
simulation is needed. Such calculations as a rule require excessive 
computing time since number of secondary particles grows exponentially 
in the cascade.
In the Monte Carlo code used in this work to speed up computations 
 we utilize, following Ref.~\cite{Kachelriess:2011bi}, the weighted  sampling 
of the cascade development. It allows us reduce the number of secondary particles.

While IC scattering occurs mostly on CMB, 
the $e^+e^-$ pair production, when it is below threshold on CMB, 
takes place on infra-red and optical components of EBL which is not 
precisely known. A number of different models have been proposed for 
EBL~\cite{EBL,Franceschini:2008tp,Stecker:2005qs,Kneiske:2003tx,
Kneiske:2010pt,Stecker:2012ta,Inoue:2012bk}. There are some upper bounds 
on EBL in the literature that were based on observations of distant blazars. 
These limits are derived without taking into account the contribution of 
cosmic rays and therefore these bounds can be relaxed~\cite{Essey:2010er}, and 
only limits based on GRBs~\cite{Abdo:2010kz} remain unaffected. The 
model of Ref.~\cite{Stecker:2005qs} is disfavoured by Fermi LAT observation 
of the photons from GRB 090902B and GRB 080916C. To obtain the range
of $\omega_{cas}$ in this work we use the baseline model from the recent 
work~\cite{Inoue:2012bk} which includes estimates of EBL for redshifts 
$z\leq 10$. In addition we use for comparison the ''best fit'' and ''lower-limit'' models of 
Refs.~\cite{Kneiske:2003tx,Kneiske:2010pt}, providing estimates of EBL 
for $z \leq 5$. The EBL model of Ref.~\cite{Franceschini:2008tp} for 
$z\leq 2$ is used only for comparison of our numerical calculations with 
work~\cite{Taylor:2011bn}.

Another poorly known factor which is crucial for consideration of the  
electromagnetic cascades from individual sources is intergalactic magnetic 
field (IGMF). Even in presence of relatively weak IGMF the angular size 
of sources can increase due to deflection of electrons and positrons 
moving along the curved trajectories with curvature  radius $R_c$
\begin{equation}
R_c=\frac{E_e}{eB}\simeq 1.1\left(\frac{E_e}{1\mbox{ TeV}}\right)
\left(\frac{B_{\perp}}{10^{-15}\mbox{ G}}\right)^{-1}\mbox{ Mpc}\,.
\end{equation}
After traversing distance $L$ the misalignment of the electron
direction with the primary $\gamma$-ray direction is given by angle $\delta$: 
\begin{equation}
\delta\simeq
\left\{
\begin{array}{ll}
\frac{\displaystyle L}{\displaystyle R_c}, & L \ll \lambda_B  \\
&\\
\frac{\sqrt{\displaystyle L\lambda_B}}{\displaystyle R_c}, & L \gg \lambda_B \,. \\
\end{array}
\right.
\label{e5:deflection}
\end{equation}
where $\lambda_B$ is IGMF correlation length. In the second case above,
many stochastic deflections were taken into account.
The deflections in the cascade can not be neglected as soon as electron 
energy-loss length becomes comparable with defocusing length i.e. the travel 
path at which electrons are deflected by maximal angle
$\delta$. The definition of $\delta$ varies for different problems. It
may  be related to experimental angular resolution or average angular 
distance between the sources. The defocusing lengths for $\delta=10^{\circ}$ 
and a range of IGMF strengths (assuming $\lambda_B \gg L$) are shown in 
Fig.~\ref{rates} (black dashed lines) together with the energy-loss length 
(red curve) for comparison. As example in case $B=10^{-15}G$ and 
$\delta=10^{\circ}$ one can infer that deflections become important for 
electron energy $E_{\rm e}\lsim 1 {\rm TeV}$ which corresponds to typical 
recoil photon energy of $3 {\rm GeV}$. Below this energy 
$\gamma$-ray flux is essentially isotropized.

Current theoretical and observational constraints on the IGMF mean  
value and correlation length are summarized in the 
review~\cite{DurrerNeronov} as
\begin{eqnarray}
10^{-17}G \lsim & B & \lsim 10^{-9}G,\label{limitB} \\
\lambda_{B} & \gsim & 1pc;  \label{limitLcor}
\end{eqnarray}
where the obtained lower limit on IGMF is based on the simultaneous observation 
of GeV and TeV gamma-radiation from the hard-spectrum blazars 
RGB J0710+591, 1ES 0229+200, and 1ES 1218+304 (Fermi/LAT-observations
in GeV, and Veritas, MAGIC and HESS observations in TeV)~\cite{Taylor:2011bn}.

In the special case of IGMF being close to it's upper limit 
$B \simeq 10^{-9}$ and for the sources emitting ultra-high energy photons or 
electrons with $E \gsim 10^{19}{\rm eV}$, the electron synchrotron losses 
should be taken into account. Pink line on Fig.~\ref{rates} represents 
energy loss-length for this process given by~\cite{jackson}. 
\begin{equation}
L^{-1}_{syn} = \frac{1}{E_{\rm e}} \frac{dE_{\rm e}}{dt} = 
- \frac{4}{3}\sigma_T \frac{B^2}{8\pi} \frac{E_{\rm e}}{m^2_{\rm e}}
\label{synchrotronAttLength}
\end{equation}
where $\sigma_T$ is the Thomson cross section, and $m_e$ is the electron 
mass. In this paper we disregard the synchrotron energy losses if not 
otherwise stated.
\begin{figure}[t]
\begin{center}
\includegraphics[angle=0,width=0.5\textwidth]{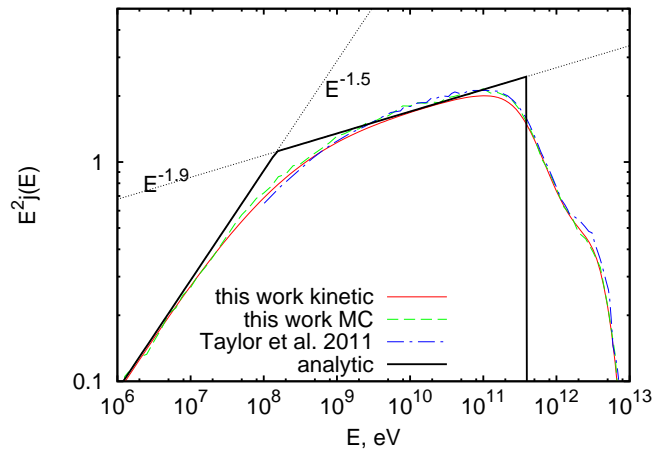}
\end{center}
\caption{Average cascade spectrum obtained using kinetic equation 
code (solid curve) and Monte Carlo code (dashed curve)  
compared with analytic calculation and with results of Monte Carlo 
simulation~ \cite{Taylor:2011bn} (dot-dashed curve), all for 
injection of $10^{14}~eV$ photons at $z=0.13$.} 
\label{compare}
\end{figure}
%%%%%%%%%%%%%%%%%%%%%%%%%%%%%%%%%%%%%%%%%%%%%%%%%%%%%%%%%%%%%%%%%
\subsection{Comparison}
\label{subsec:comparison}
In this section we compare the cascade spectra obtained using the 
kinetic equation and Monte Carlo codes from the present work with 
other numerical simulations and with analytic calculations. 
Fig.~\ref{compare} demonstrates level of agreement between our
numerical calculations, simulation~ \cite{Taylor:2011bn}, and analytic
calculation for a source at $z=0.13$ injecting $10^{14}eV$ 
photons and using EBL model of Ref.~\cite{Franceschini:2008tp}
for photon absorption. 
One may see in this figure the familiar two cascade energy spectra 
$\propto E^{-1.5}$ and $\propto E^{-1.9}$ and also the steeper high 
energy feature in a good agreement in all three numerical simulations. 
As will be demonstrated below this feature plays an important role 
in determination of the cascade energy density $\omega_{\rm cas}$.

The diffuse gamma-ray spectrum as presented by Fermi LAT in 2015  
\cite{Fermi2014} shows the steepening which starts at 
$1\times 10^{11}$~eV and continues as more sharp cutoff at 
$\epsilon = 2.5 \times 10^{11}$~eV. The both high energy features 
in theoretical spectra in Fig.~\ref{compare}, $\propto E^{-1.9}$
and more steep feature above it, are more flat than that 
observed by Fermi LAT and it results  in the upper limit on 
$\omega_{\rm cas}$. The reason may be easily understood from  
Fig.~\ref{compare}. 

For given large enough $\omega_{\rm cas}$ the realistic, numerically 
calculated, spectra from Fig.~\ref{compare} can intersect the Fermi 
high energy tail at some energy. It means that at energy above the 
crossing the calculated  cascade flux is larger than the measured
Fermi flux. To eliminate this contradiction one must lower the
calculated cascade flux, i.e. $\omega_{\rm cas}$. This procedure 
results in the upper limit on $\omega_{\rm cas}$.

The case of analytic spectrum in Fig.~\ref{compare} is quite different. 
It has a sharp cutoff at $\epsilon_\gamma \approx 400$~GeV close to Fermi 
cutoff energy $\epsilon \approx 250$~GeV and thus quite larger  
$\omega_{\rm cas}$ is allowed. 

All effects which make lower the high-energy 
cutoff in the calculated cascade spectrum, e.g. high red-shift of production, 
{\em increase} the upper limit on $\omega_{\max}$.

%%%%%%%%%%%%%%%%%%%%%%%%%%%%%%%%%%%%%%%%%%%%%%%%%%%%%%%%%%%%%%%%%%
\section{Universality of the cascade spectra in numerical simulations}
\label{sec:universality}
\begin{figure}
        \centering
        \begin{subfigure}[b]{0.5\textwidth}
                \includegraphics[width=\textwidth]{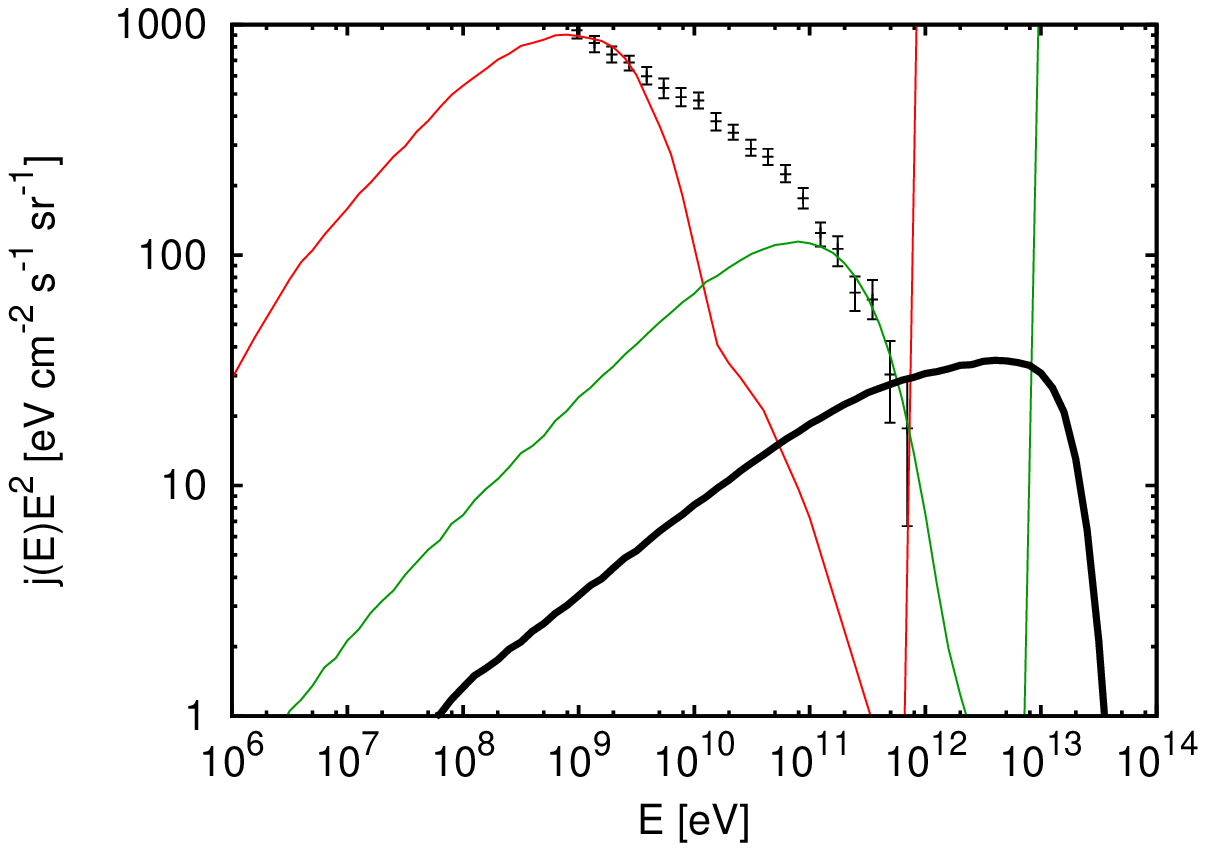}
                \caption{%diffuse flux 
                $z_{s}=0.01$}
                \label{univE:z0.01}
        \end{subfigure}%
        \quad
         ~%add desired spacing between images, e. g. ~, \quad, \qquad etc.
          %(or a blank line to force the subfigure onto a new line)
        \begin{subfigure}[b]{0.5\textwidth}
                \includegraphics[width=\textwidth]{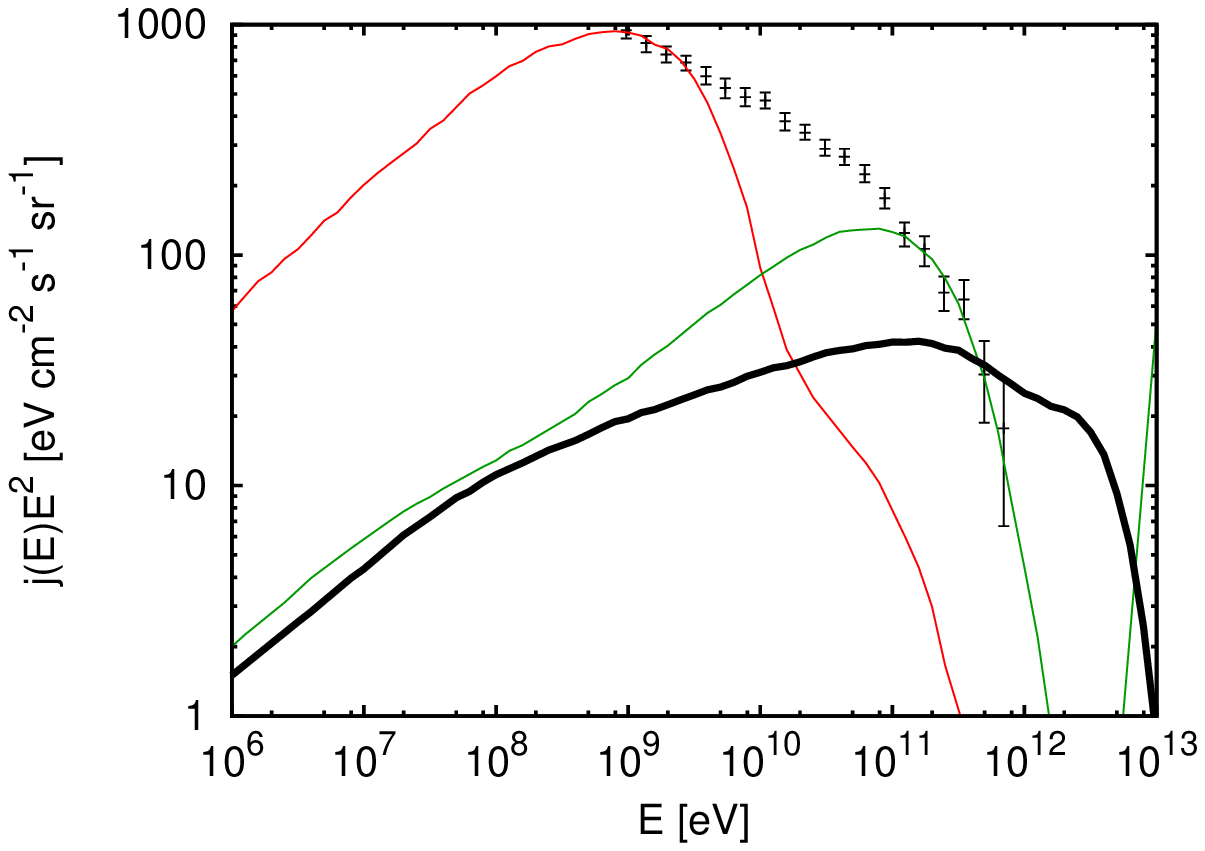}
                \caption{%diffuse flux 
                $z_{s}=0.1$}
                \label{univE:z0.1}
        \end{subfigure}
        \quad
        \begin{subfigure}[b]{0.5\textwidth}
                \includegraphics[width=\textwidth]{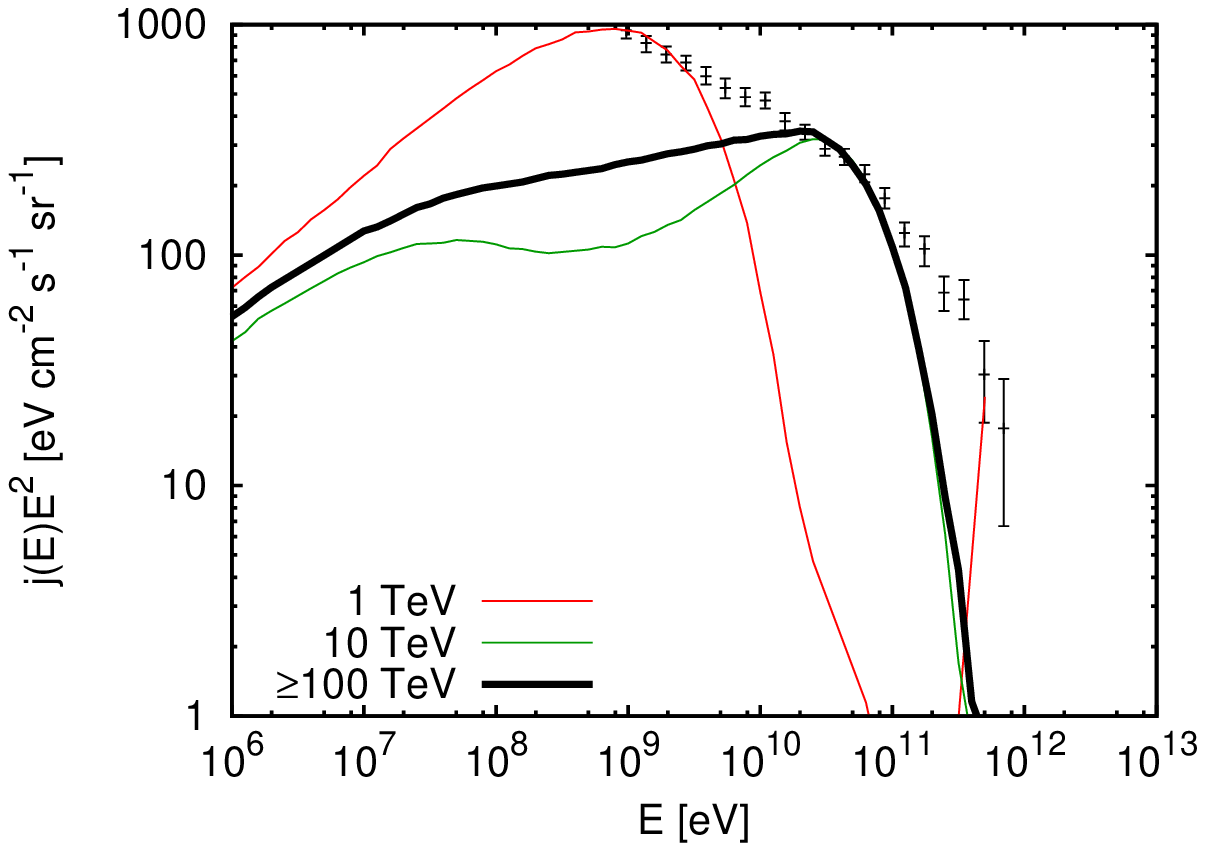}
                \caption{%diffuse flux 
                $z_{s}=1$}
                \label{univE:z1}
        \end{subfigure}
         
        \caption{Diffuse cascade spectrum
        from sources injecting photons with energy 
$E_{s}=1~{\rm TeV}$ (red curves), $10~{\rm TeV}$ (green curves) and $E_{s} \geq 100~{\rm TeV}$ (black thick curves) 
at redshifts $z=0.01$,~ $z=0.1$,~ $z=1$. 
Photon fluxes are 
limited from above by the Fermi EGB flux, shown by black errorbars,
with foreground model B~\cite{Fermi2014}. 
The curves obtained for initial photon energies greater than $100~{\rm TeV}$ coincide 
in all figures. In calculations the EBL flux from Ref.~\cite{Inoue:2012bk} 
is used.}
\label{univE}
\end{figure}

Universality of the cascade energy spectrum was discovered first in 
analytic calculations~\cite{book}, and we will start our discussion
from analytic dichromatic model of section \ref{sec:universal}. This
universality is clearly seen from  Eq.~(\ref{eq:gamma-spectrum}), 
where two joint energy spectra $\propto E^{-3/2}$ and $\propto E^{-2}$ 
appear, divided by two boundary energies $\calE_X$ and $\calE_\gamma$, 
built from basic parameters of dichromatic model $\eps_{\rm cmb}$ and   
$\eps_{\rm ebl}$. The main features of this universal 
spectrum include: (i) The same energy shape of the spectrum produced 
by initial photon/electron if its primary energy is sufficiently high. 
Ultimately the scale can be as high as 
$E_0 \gsim m_e^2/\eps_{\rm cmb} \sim 0.4$~PeV and it is the basic one
in analytic calculations. The lower universality scale 
$E_0 \sim 100$~TeV is found in numerical simulations and this result 
is very natural and may be expected apriori.  In analytic approach the
cutoff energy $\calE_\gamma^{\rm cmb}=0.4$~PeV is consequence of 
monochromatic spectrum of CMB photons 
$\epsilon_{\rm cmb}=6.3\times 10^{-4}$~eV. In numerical simulations 
absorption occurs on high-energy tail of Planckian distribution of CMB
photons and absorption produced at $\calE_\gamma^{\rm cmb} \approx 100$~~
TeV when $\ell_{\rm abs}$ reaches $c/H_0$. 
(ii) The cascade energy spectrum is the same for any injection spectrum 
$Q(E_s)$ at $E_s \gsim E_0$ (in other words cascade spectrum forgets what 
injection spectrum produced it). (iii) The cascade spectrum does not depend 
on distance to the point where cascade started, and (iv) Energy density of 
the cascade $\omega_{\rm cas}$ is the only cascade characteristic which 
determines the spectrum. 

These properties of 'analytic' cascades will be referred to as {\em strong 
universality}. 

For realistic cascades in the expanding universe the strong universality, 
as it is formulated above, is not valid. The cascade spectrum observed
at $z=0$ depends on redshift of production $z_s$, e.g. due to dependence 
$\calE_X$ and $\calE_\gamma$ on $z_s$ and simply due to redshift of the 
spectrum. Consider for example the generation rate of photons/electrons 
in expanding universe in the form $Q(E_s,z_s)=\phi (E_s) R(z_s)$ with $E_s$ 
above the universality scale $E_0$. All cascades produced at fixed redshift
$z_s$ have the same cascade spectrum $q_{\rm cas}(E)$ at $z=0$. However,  
cascades originated at different $z_s$ have the different cascade spectra at 
$z=0$, and integration over $z_s$ results in cascade spectra, which depend  
on distribution of production rate $Q(E_s,z_s)$ over $z_s$. Predicted cascade 
spectrum is determined not only by $\omega_{\rm cas}$ but e.g. by parameters 
of cosmological evolution of the sources. Universality of the cascade
spectra remains only for a subclass of the sources with approximately equal 
redshifts.  In summary, the sources in expanding universe with the
generation rate  $Q(E_s,z)$, with $E_s$ above the universality
scale $E_0$ and with arbitrary dependence on $z$ (e.g. evolution 
$(1+z)^m$ up to $z_{\max}$) have the following properties of universality:
(i) energy shape of em-cascade at $z=0$ is independent from $E_s$ or from 
injection spectrum $\phi (E_s)$, i.e. for fixed $z_s$ the shape is the
same for different $E_s$, or for different spectra $\phi (E_s)$, (ii) cascade 
spectrum at $z=0$ is not uniquely determined by $\omega_{\rm cas}$, but 
depends also on evolution of $Q(E,z)$ with $z$, e.g. through parameters 
$m$ and $z_{\max}$, (iii) a subclass of the sources with the same (or 
almost the same) redshift $z_s$ has the identical cascade spectrum at $z=0$, 
which is characterized by the single parameter $\omega_{\rm cas}$. 

This case can be referred to as {\em weak universality}. 

Class of sources with the fixed redshift of production $z_s$, i.e  with
production rate $Q(E_s,z)=\phi(E_s)\delta(z-z_s)$ has all properties of 
strong universality, except (iii), and can be attributed to 
{\em strong universality} for arbitrary generation spectrum $\phi(E_s)$ with 
$E_s$ above the universal scale $E_0$. The corresponding cascade spectrum 
is characterized by a single parameter $\omega_{\rm cas}$ for any fixed $z_s$.

Consider $z$-fixed sources in some details, starting from low initial
energy $E_s < E_0$. At increasing  $E_s$ the $z=0$ cascade spectrum 
must evolve to strongly-universal spectrum and reaching it at scale $E_0$ 
stay unchanged. Fig.~\ref{univE} illustrates this statement. The figure 
presents the diffuse fluxes generated  by the population of sources with 
the fixed energy of injected photon $E_s$ and with fixed redshift $z_s$.
The diffuse spectrum {\em shape} at $z=0$ doesn't depend on $E_s$
as long as $E_s\gsim\calE_{\gamma}^{cmb}$.
In fact, for remote enough sources at $z\gsim0.1$, universality is
reached at the scale $E_0 \gsim 100~{\rm TeV}$. This observation  was 
tested for all EBL models used in this work.

%%%%%%%%%% discussion of diffuse spectra in Fig. 9  %%%%%%%%%
The numerically calculated diffuse cascade spectra from the sources 
with fixed redshift $z_s$ and energy $E_s$ are shown in 
Figs~\ref{univE:z0.01} - \ref{univE:z1}. The spectra are expected to be 
strongly universal  when energy $E_s$  exceeds the largest scale 
$E_0=\calE_\gamma^{\rm cmb}=0.4$~PeV. All three figures (a), (b) 
and (c) demonstrate that the spectra are universal at $E \geq E_0$, 
where $E_0=100$~TeV.
This energy can be considered as energy scale of universality in 
numerical simulations (see discussion above).     
Apart from identical spectra they have at large $z_s \geq 0.1$ the
predicted standard spectrum $\propto E^{-3/2}$ at low energy, 
$\propto E^{-1.9}$, at intermediate energies with highest energy 
feature in the end of the spectrum (cf with Fig.~\ref{compare}).  
The spectra with $E_s = 100$~TeV (black thick curves) and with all 
energies $E_s \gsim 100$~TeV are indistinguishable in Figs~\ref{univE}.

The cascade universality makes it hard to extract the injection spectrum  
from the diffuse spectrum observed at $z=0$. On the other hand
the comparison of the calculated and observed spectra allows us to
estimate the upper limit on the main characteristic of the universal
cascade spectrum, the cascade energy density  $\omega_{\rm cas}$ at 
$z=0$. This problem will be considered in the next section.

%%%%%%%%%%%%%%%%%%%%%%%%%%%%%%%%%%%%%%%%%%%%%%%%%%%%%%%%%%%%%%%%%%

\section{Upper limit on $\omega_{\rm cas}$ from Fermi LAT data}
\label{sec:omega}

The upper limit on the cascade energy density given by 
Eq.~(\ref{eq:omega2011}) as $5.8\times 10^{-7}$~eV/cm$^3$ has been 
derived in \cite{BGKO} from the first-year Fermi data \cite{FermiLAT}.   
Here we obtain this limit from 50 months Fermi LAT observations 
\cite{Fermi2014} using somewhat different approach. 

Fermi LAT~\cite{FermiLAT} presents two kinds of extragalactic gamma-ray fluxes 
in energy range 100~MeV - 820~GeV: EGB (Extragalactic $\gamma$-ray Background) 
and IGRB (Isotropic diffuse $\gamma$-ray Background), 
 EGB presents the total extragalactic gamma-ray flux, from which about half is given by the resolved 
individual sources. In the cascade calculations we use these fluxes to
normalize the upper limits on the cascade energy density $\omega_{cas}$.
The highest and most conservative upper limit on theoretical energy 
density $\omega_{cas}$ is imposed by Fermi EGB flux
and is marked as $\omega_{\rm cas}^{\rm tot}$. In isotropic case  
the estimate of $\omega_{\rm cas}^{\rm iso}$ can be obtained from Fermi IGRB 
flux \cite{Fermi2014} assuming additionally  highly homogeneous 
distribution of gamma-ray sources or astrophysical $\gamma$-ray generation 
scenarios where Intergalactic Magnetic Field (IGMF) must be high enough to 
isotropise the cascade electrons and positrons (or their parent particles) 
in the space.

EGB flux is higher than IGRB, and the two calculated values of 
$\omega_{\rm cas}$ follow this hierarchy. Both fluxes are described by 
power-law spectrum with index $\gamma \approx 2.3$ and with more steep 
highest energy tail starting at $E_{\rm cut}=\epsilon\approx 250$~GeV 
(beginning of the steepening). The nature of gamma-ray flux above 
$E_{\rm cut}$ is not well known, but we consider it as diffuse flux. 
This high-energy steep component (tail) is responsible for stronger upper
limit on $\omega_{\rm cas}$ in comparison with earlier paper~\cite{BGKO2011}.
The realistic, numerically calculated, spectra in Fig.~\ref{compare} 
are flatter than the highest-energy Fermi tail and  can intersect it at 
some energy (see section~\ref{subsec:comparison}).

%%%%%%%%%%%%%%%%%%%%%%%%%%%%%%%%%%%%%%%%%%%%%%%%%%%%%%%%%%%%%%%%%%%%%%%%%%%%%%
\subsection{Upper limit on $\omega_{\rm cas}$ in analytic calculations}
\label{sec:omega-analytic}

To illustrate our method of calculating  $\omega_{\rm cas}$ we will 
consider first a simple example of analytic model followed then by the 
accurate numerical calculations. 

Consider the energy of Fermi IGRB spectrum cutoff 
$\epsilon = 2.5\times 10^5$~MeV, where the measured flux is 
$J_\gamma^{\rm igrb}=4.80\times 10^{-16}cm^{-2}s^{-1}sr^{-1}MeV^{-1}$.
We use here the analytic dichromatic model with 
$\eps_{\rm cmb}=6.3\times 10^{-4}$~eV and $\eps_{\rm ebl}\approx 1$~eV,
which provides the high-energy cutoff in the cascade spectrum 
$\calE_\gamma=m_e^2/\eps_{\rm ebl}=2.61\times 10^{5}$~MeV, practically
the same as observed in Fermi IGRB spectrum $\epsilon=E_{\rm cut}$.
We take the cascade spectrum as $J_{\rm cas}(E) \propto E^{-1.9}$
at $\calE_X \leq E \leq \calE_\gamma$ and $E \propto E^{-1.5}$ at 
$E \leq \calE_X$. Then the cascade energy density can be calculated 
using the cascade flux $J_{\rm cas}(E)$ as  
\be
\omega_{\rm cas}=\frac{4\pi}{c}\left (\int_0^{\calE_X} dE E J_{\rm cas}(E)
+\int_{\calE_X}^{\calE_\gamma} dE E J_{\rm cas}(E) \right ) 
\label{eq:illustrOmega_cas}
\ee
The most restrictive relation we use in calculation of 
Eq.~(\ref{eq:illustrOmega_cas}) is given by the cascade flux 
in energy interval $\calE_X \leq E \leq \calE_\gamma$  
\be
J_{\rm cas}(E)=J_{\rm igrb}(\epsilon) (E/\epsilon)^{-1.9},
\label{eq:illustr-norm}
\ee
which includes normalization of the cascade flux by 
Fermi IGRB flux $J_{\rm igrb}$ at energy $\epsilon=E_{\rm cut}$.
This particular condition provides the low upper limit 
on $\omega_{\rm cas}$ in this estimate. 

For interval $E \leq \calE_X$ we use 
\be
J_{\rm cas}(E)=J_{\rm igrb}(\epsilon)(\calE_X/\epsilon)^{-1.9}
(E/\calE_X)^{-1.5}
\label{eq:-3/2}
\ee
As a result we obtain the upper limit on energy density of cascade 
radiation  
\be
\omega_{\rm cas} \leq 6.6 \frac{4\pi}{c} \epsilon^2 J_{\rm igrb}(\epsilon)
=8.3 \times 10^{-8}~ {\rm eV/cm}^3,
\label{eq:illustrOmega_casN}
\ee
to be compared with much larger cascade upper limit 
$\omega_{\rm cas}^{\max}=5.8\times 10^{-7}$~eV/cm$^3$
obtained in \cite{BGKO2011}  and used in \cite{Ahlers2010}.
The reason is that for normalization of calculated flux 
we used the measured flux at energy $\epsilon=2.5 \times 10^5$~MeV,
which is located below the Fermi $E^{-2.3}$ approximation of the flux.  
However, this argument implies the further decrease of 
$\omega_{\rm cas}^{\max}$. In more realistic numerical simulations 
there is the high-energy tail (see Fig.~\ref{compare}) and
intersection of this tail with steep high-energy Fermi IGRB tail 
demands lowering the calculated flux, i.e. further suppression
of $\omega_{\rm cas}$. Another reason of modification of 
calculated spectrum is connected with interpretation of two 
highest energy points in the Fermi LAT spectrum above 
$2.5\times 10^5$~MeV. If these points belong to isotropic diffuse 
radiation the high-energy theoretical tail must be shifted downward 
and it results in lowering of $\omega_{\rm cas}$. However, in case  
the weak universality $\omega_{\rm cas}$ is not the only parameter 
which influence the flux, it could be that other parameters,
e.g. the cosmological evolution of the sources, can shift the flux
upward at the same $\omega_{\rm cas}$.

The stronger upper limit on $\omega_{\rm cas}^{\rm iso}$ obtained 
here put the stronger upper limit on the flux of UHE
extragalactic protons and cosmogenic neutrinos, in comparison with
\cite{BGKO2011,Ahlers2010,GKS2012}. The effect of increasing 
the fraction  of resolved sources~\cite{DiMauro2016} diminishes 
further $J_{\rm cas}(E)$ and $\omega_{\rm cas}$ given above.

%%%%%%%%%%%%%%%%%%%%%%%%%%%%%%%%%%%%%%%%%%%%%%%%%%%%%%%%%%%%%%%%%%

\subsection{Upper limit on $\omega_{\rm cas}$ in numerical simulations}
\label{sec:omega-numerical}

Now we will proceed to consideration of $\omega_{\rm cas}$ using the 
calculation of the cascade spectra in numerical simulations, kinetic
equations and Monte Carlo. These calculations provide us
with the shape of the cascade spectra at $E < \epsilon _\gamma$, and 
normalization by 
the observed Fermi LAT spectrum allows us to obtain the values of 
$\omega_{\rm cas}$, the final aim of our research in this paper. 

Below we calculate $\omega_{cas}^{tot}$ and $\omega_{cas}^{iso}$ for 
two cases: 
(i) the redshift-fixed photon-electron sources with very high energy 
$E_s$ and (ii) redshift-distributed sources with 
injection rate of photons/electrons $Q(E_s,z)$ smoothly dependent on 
$z$. The case (i) results in strong universality and (ii) for general
case of $z$ dependence, in weak universality. The scale of very high 
energy is given by $E_s\gsim\calE_{\gamma}^{cmb} \gsim 0.4$~PeV for
analytic calculations, though as argued above and as 
Figs~\ref{univE}a - \ref{univE}c show,  
the universal shape of the spectra is reached already at energy scale 
$E_s \sim 100$~TeV.

Let us now come over to numerical calculation of the cascade spectra
and to evaluation of $\omega_{\rm cas}$. We consider two cases:\\*[2mm]

(i)~~ The redshift-fixed and energy-fixed sources with injection rate
\begin{equation}
Q(E,z) \propto \delta(z-z_s) \delta(E-E_s),
\label{delta_source}
\end{equation}
and\\*[2mm]
(ii)~~ Redshift-distributed sources with injection rate 
\begin{equation}
Q(E,z) = (1+z)^{3+m} \delta(E - E_s),~ {\rm at}~ z<z_{\max},
\label{cont_source}
\end{equation}
with $1<z_{max} < 5$ and $0<m<5$  (the case $m=0$ corresponds to 
constant source density in comoving frame).

In calculations for both cases we will keep $E_s \geq 100$~TeV 
to provide universality.\\

The calculated spectra will be normalized by Fermi LAT spectra, EGB and 
IGRB. The highest and most conservative upper limit on energy density 
$\omega_{\rm cas}$ is imposed directly by EGB flux, which includes also
the flux of resolved extragalactic discrete sources. To be even more
conservative we take the fluxes of EGB (and IGRB too) as maximal one 
allowed by systematic uncertainties.  Using these two fluxes we obtain 
the upper limits on $\omega_{\rm cas}^{\rm igrb}$ and         
$\omega_{\rm cas}^{\rm egb}$, the latter will be considered as maximally
allowed  $\omega_{\max}$. 

In case (i) the spectra are shown for three values of $z_s$ in 
Figs.~\ref{univE:z0.01} - \ref{univE:z1}. At energies $E_s$ higher than 
100~TeV all spectra are the same. 
\begin{figure}[t]
\begin{center}
\includegraphics[angle=0,width=0.5\textwidth]{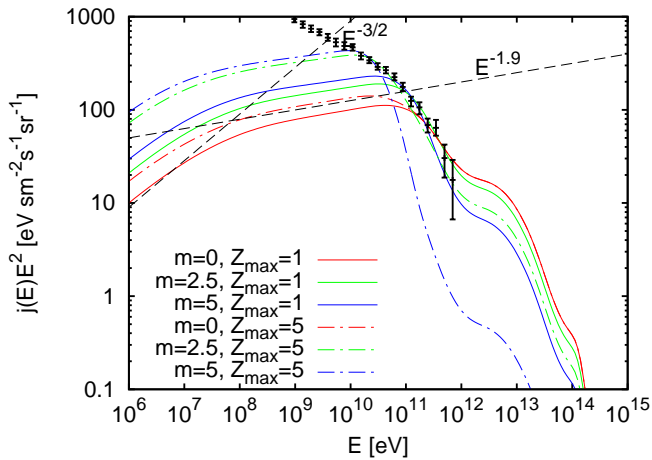}
\end{center}
\caption{Cascade spectra for source distribution~(\ref{cont_source}), 
with $E_s=1$~PeV, with various $m$ and $z_{max}$, and with EBL model of 
Ref.~\cite{Kneiske:2003tx}. The spectra are normalized by 
Fermi LAT EGB flux (with foreground model B)~\cite{Fermi2014} 
shown by black errorbars.
} \label{homo_spec}
\end{figure}
Fig.~\ref{homo_spec} presents the cascade spectra  for continuous
source distribution (\ref{cont_source}), which illustrates the
dependence of the cascade spectra on evolution parameters $m$ and 
$z_{\max}$. 

From Fig.~\ref{homo_spec} one may observe that spectra with low 
$z_{\max}=1$ have large cutoff-energy $E_{\max}$ and to avoid the
contradiction with Fermi data one must shift the calculated spectrum 
downwards, diminishing thus $\omega_{\rm cas}$.

Spectra with large $z_{\max}=5$ have lower $E_{\max}$ due to redshift 
factor $(1+z)$ and respectively larger $\omega_{\rm cas}$. Dependence on 
$m$ works in the similar way.  Fig.~\ref{homo_spec} allows us to 
calculate $\omega_{\rm cas}$ using the corresponding curves. 
As a result the constraints on energy density of cascades with 
large $m$ and $z_{max}$ are relaxed.

We already have seen such effect in section~\ref{sec:comparison} for 
the case of analytic solution with the sharp high-energy cutoff,
which starts at low energy (see Fig.~\ref{compare})
Now we can generalize the both cases formulating what will be called  
{\em ``$E_{\max}$ rule''} used here and below. It 
reads: {\em Increasing $E_{\max}$ in the calculated cascade spectrum
suppresses $\omega_{\rm cas}$}. Indeed, Fig.~\ref{homo_spec} 
shows that increasing $E_{\max}$ needs lowering 
the total curve $J(E)$ to avoid the excess of predicted flux     
$J(E)$ over the observed Fermi flux. For changing $E_{\max}$ one 
may use, for example, the cosmological evolution: increasing 
$z_{\max}$ results in decreasing $E_{\max}$ at $z=0$ by factor 
$(1+z)$. 

Another feature observed in Fig.~\ref{homo_spec} is the standard 
energy spectra $\propto E^{-1.5}$ transforming to $\propto E^{-1.9}$ 
at higher energy (see these curves in the figure).

        \begin{figure}[t]
        %\centering
        \begin{subfigure}[a]{0.5\textwidth}
                \includegraphics[width=\textwidth]{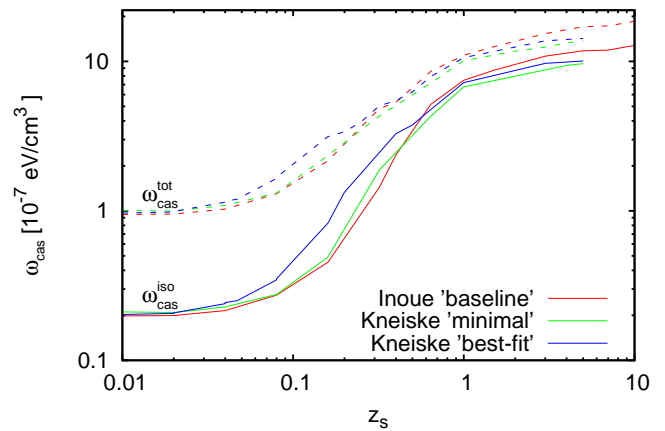}
                \caption{sources at fixed $z$}
                \label{omega:point}
        \end{subfigure}%
        \quad
         ~%add desired spacing between images, e. g. ~, \quad, \qquad etc.
          %(or a blank line to force the subfigure onto a new line)
        \begin{subfigure}[b]{0.5\textwidth}
                \includegraphics[width=\textwidth]{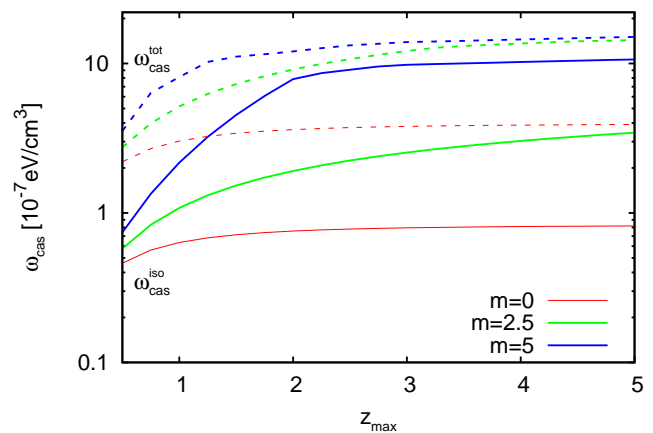}
                \caption{sources distributed in $z$}
                \label{omega:cont}
        \end{subfigure}
        \caption{Upper limits on the cascade energy density 
$\omega_{\rm cas}$ obtained for source distribution with 
fixed $z_s$, Eq.~(\ref{delta_source} (panel a), and with continuous $z_s$ 
distribution, Eq.~\ref{cont_source} (panel b). The total upper limits 
$\omega_{\rm cas}^{\rm tot}$ are obtained using Fermi EGB flux (dashed 
lines) and isotropic upper limits $\omega_{\rm cas}^{\rm iso}$ - 
using Fermi IGRB flux (full lines). The various EBL models are used in
simulations shown in panel (a): Ref.~\cite{Inoue:2012bk} (red lines), 
Ref.~\cite{Kneiske:2003tx} 
(blue lines), and Ref.~\cite{Kneiske:2010pt} (green lines). In the lower panel 
(b) EBL model of Ref.~\cite{Inoue:2012bk} is used. }
\label{omega}
\end{figure}

%%%%%%%%%%%%%%%%%%%%%%%%%%%%%%%%%%%%%%%%%%%%%%%%%%%%%%%%%%%%%%%%%%%%%%%%%%

In Fig.~\ref{omega} we present the maximum cascade energy density 
$\omega_{cas}^{iso}$ consistent with Fermi IGRB flux and $\omega_{cas}^{tot}$ 
consistent with Fermi EGB flux~\cite{Fermi2014} for two cases of
source distribution~(\ref{delta_source}) and~(\ref{cont_source}). To 
obtain these quantities we normalize the cascade spectrum calculated 
in numerical simulations by IGRB or EGB fluxes~\cite{Fermi2014}. 

In the  panel (a) we show the energy density $\omega_{cas}^{tot}$ 
and $\omega_{cas}^{iso}$ calculated in the model  with fixed redshift
of the source $z_s$ and with fixed energy $E_s$ of primary photon/electron. 
This is the case of strong universality, when the cascade spectrum 
is determined  by single parameter $\omega_{\rm cas}$.  Uncertainties, 
caused by different EBL models are not large, and values of 
$\omega_{cas}^{tot}$ are larger than $\omega_{cas}^{iso}$, as expected. 
The values of $\omega_{\rm cas}(z)$ at large $z$ exceeds that at small
$z$ according to $E_{\max}$ rule: large $z$ gives small $E_{\max}$, small 
$E_{\max}$ results in large flux, and hence in large $\omega_{\rm  cas}$. 
One can see this effect in Fig.~\ref{omega:point}. 

In Fig.~\ref{omega:cont} the case of more realistic continuous 
$z$-distribution, as given by Eq.(\ref{cont_source}), is presented. 
It is described by weak universality, when cascade spectrum depends. 
apart from $\omega_{\rm cas}$, on other parameters, in particular 
on parameters of cosmological evolution $m$ and $z_{\max}$. 
The low $z$ and large $z$ regimes exist here, too, being provided by
$E_{\max}$ rule.  

The lowest $\omega_{\rm cas}^{\rm iso}=4\times 10^{-8}$~eV/cm$^3$,
is obtained for $z_{\rm max} \lsim 1$ and for absence of evolution $m=0$.
It is an order of magnitude lower than the limit  $5.8\times 10^{-7}$~ eV/cm$^3$ found in~\cite{BGKO2011}
for the secondary photons produced during propagation of UHECR. The limit appears very restrictive for fluxes
of protons in UHECR and cosmogenic neutrinos produced at $z_{\rm max} \lsim 1$ and in absence of evolution $m=0$. 

Non-evolutionary models with m=0 and $z_{\max} \gsim 2$ have 
$\omega_{\rm cas}^{\rm iso} \leq 8\times 10^{-8}$~eV/cm$^3$ (see 
Fig.~\ref{omega} panel b) which allows some of non-evolutionary 
UHECR proton models from Table 1 of~\cite{BGKO}. The models with strong 
evolution $m=5$ and $z_{\max} \gsim 2$ allows large 
$\omega_{\rm cas}^{\rm iso} \approx 8\times 10^{-7}$~eV/cm$^3$ favourable 
for UHECR proton models with strong evolution and large $z_{\max}$.

The smallness of $\omega_{\rm cas}$ produced in cosmic ray models 
in comparison with $\omega_{\rm cas}^{\rm iso}$ measured by Fermi LAT
is not the only criterion for successful cosmic ray model. It must
satisfy another more sensitive criterion: {\em not to exceed the Fermi IGRB 
flux in the highest-energy bin} (in fact this criterion enters the
comparison of energy densities as integral characteristic).
Unfortunately the IGRB flux estimate in the highest energy bin
is strongly model dependent and suffers from low statistics.
We have considered problem of survival of UHECR proton models in details in the 
separate work with emphasizing the role of the highest energy bin (in preparation).
This problem was  already studied in the works~\cite{Ahlers2010,BGKO2011,GKS2012}
and most recently in different approach in Ref.~\cite{Liu:2016brs}.

The  above analysis is based on EGB and IGRB fluxes derived using 50 months of Fermi-LAT 
observation~\cite{Fermi2014}. The two recent catalogs of
observations which appear later~\cite{Fermi2015} and~\cite{Fermi2015a}
probably will change further our results. These catalogs are based on the new program of  analysis, 
Pass 8, with the improved  reconstruction and classification of
events, and  with the time of observation increased to 6 years.  The
analysis of the work~\cite{DiMauro2016} shows that the considerable fraction of  
high energy events above 50 GeV can be attributed to unresolved sources most of which are blazars. 
For EGB flux, contribution of blazars according to this work reaches 
$86^{+16}_{-14}\%$. This implies a stronger bound on the true isotropic flux.
This effect can be roughly described by inequality
\begin{equation}
\omega_{\rm cas}^{\rm\prime iso} \lsim 0.28 \omega_{\rm cas}^{\rm tot}
\end{equation}

%%%%%%%%%%%%%%%%%%%%%%%%%%%%%%%%%%%%%%%%%%%%%%%%%%%%%%%%%%%%%%%%%%%%%%%
\section{Summary}
\label{sec:conclusion}
Using both analytic approach and numerical simulations we have described 
the development of electromagnetic cascades in the universe in
presence of CMB and EBL background radiations. The cascades develop 
due to IC scattering on most numerous CMB photons 
$e+\gamma_{\rm cmb} \to e'+\gamma'$ and pair-production on less
numerous EBL photons $\gamma+ \gamma_{\rm ebl} \to e^-e^+$. A 
primary particle below is called photon, though electron is implied
under this name, too.

For analytic calculations we use {\em dichromatic model} with fixed 
energy of CMB photons $\eps_{\rm cmb}=6.3\times 10^{-4}$~eV and EBL 
photons with $\eps_{\rm ebl}=0.68$~eV. In the cascade photon spectrum 
there are two characteristic energies: absorption energy 
$\calE_\gamma^{\rm ebl} \sim m_e^2/\eps_{\rm ebl}$ and Inverse Compton 
energy of a photon $\calE_X=(1/3)(\calE_\gamma/m_e)^2\eps_{\rm cmb}$,
produced by a born electron/positron in $\gamma+\gamma_{\rm ebl}$ 
collision. Thus in analytic dichromatic model we have: 
\begin{eqnarray}
\calE_\gamma^{\rm ebl}=\frac{m_e^2}{\eps_{\rm ebl}}=
3.9\times 10^{11}~{\rm eV} \nonumber\\
\calE_X= \frac{1}{3}\calE_{\gamma}^{\rm ebl}
\frac{\epsilon_{\rm cmb}}{\epsilon_{\rm ebl}} 
= 1.2\times 10^8~{\rm eV}. 
\label{Egamma,E_X}
\end{eqnarray}
The cascade initiated at large distance by very high energy 
photon/electron has spectrum  given by Eq.~(\ref{eq:gamma-spectrum}), 
which is $\propto E^{-3/2}$ in low-energy regime $E \lsim \calE_X$, 
with $\propto E^{-2}$ at intermediate energies 
$\calE_X \lsim E \lsim \calE_\gamma$, and with a high energy cutoff 
at $\calE_\gamma$, where $\calE_\gamma$ numerically can differ from  
$\calE_\gamma^{\rm ebl}$ in Eq.~(\ref{Egamma,E_X}) due to different 
values of $\epsilon_{\rm ebl}$.

The numerical simulations confirm the analytic 
spectrum with index $\gamma_1=3/2$ being exact, $\gamma_2=2$ being 
approximate ($\gamma_2 \approx 1.9$ in numerical simulations) and with 
a sharp high-energy cutoff for a source at very large distance. 
The artificially sharp energy transitions in analytic spectrum at 
$E = \calE_X$ and $E=\calE_\gamma$. appear in numerical simulations 
as continuous transition features. This is most important difference between 
analytic solution and more precise numerical simulation.

The remarkable feature found in both analytic and numerical
solutions is {\em universality} of the cascade spectrum.
In analytic solution the strong universality is seen explicitly 
from the spectrum given by Eq.(\ref{eq:gamma-spectrum}), where the shape 
of the spectrum does not depend on initial energy of the primary
photon at $E_s> E_0$, and on distance to observer $r$, unless it is too 
small. The energy scale of universality is 
$E_0=\calE_\gamma^{\rm cmb}=0.4$~PeV. The energy $\calE_X$  which separates 
two regimes, $\propto E^{-3/2}$ and 
$\propto E^{-2}$, and energy of spectrum cutoff $\calE_\gamma$    
are build from the main physical constants of the model 
(see Eq.~(\ref{eq:benchmark})) and the cascade 
energy spectrum does not depend on variables of the model, in particular 
on initial energy $E_s$ and distance to the source $r$.
The main features of this universality, specific for analytic 
model developed in section \ref{sec:universal}, are: (i) the 
same energy shape of the cascades produced by primary 
photon/electron if its energy $E_s$ is larger than 
universality scale $E_0$, (ii) the same cascade energy spectrum  
for any injection spectrum $Q(E)$ at $E \geq E_0$ (in other words 
cascade spectrum forgets what injection spectrum produced it), 
(iii) indepedence of spectrum shape from distance to the point 
where cascade started, and (iv) energy density of the cascade 
$\omega_{\rm cas}$ as the only cascade parameter which determines
the spectrum shape. 

These properties of 'analytic' cascades are referred to as 
{\em strong universality}. 

For realistic cascades in the expanding universe, property (i) is modified 
as follows: cascades initiated at the fixed redshift $z$ by 
photon/electron with sufficiently high energy $E_s > E_0$ turn at $z=0$ into 
cascades with spectrum independent of $E_s$ but dependent on  $z$. 
These spectra are calculated by numerical simulation. Property (ii) remains 
almost the same: cascade spectra initiated at the same $z$ with different 
injection spectra $Q(E)$ are almost identical. However, if injection spectrum 
$Q(E,z)$ smoothly changes with $z$ the total diffuse spectrum obtained by 
integration over $z$ is not universal; the spectra are different for different 
dependence of $Q(E,z)$ on $z$. The diffuse spectra for various $Q(E,z)$ are 
determined in this case not only by $\omega_{\rm cas}$ but also by other 
parameters from $Q(E,z)$ distribution. 

We refer to the case described above as {\em weak universality}.  Some
of the described properties of these cascades, studied in numerical 
simulations are shown in Fig.~\ref{univE}.    

The main aim of this paper is to obtain the upper limit on
the cascade energy density $\omega_{\rm cas}$ using the Fermi LAT 
flux of gamma radiation.  In the analytic model with strong 
universality, there is the direct and transparent method which 
allows us to obtain strong enough upper limit on the cascade energy 
density. This upper limit follows from Eq.~(\ref{eq:illustrOmega_cas}) 
and results in $\omega_{\rm cas} \leq 8.3\times 10^{-8} eV/cm^3$.
The specific property of this limit is the high-energy cutoff, i.e. 
$E_{\max}$, given by  $\calE_\gamma = m_e^2/\eps_{\rm ebl}$, which at 
$\eps_{\rm ebl} \sim 1$~ eV coincides with cutoff energy in 
Fermi IGRB spectrum. The limit is stronger if the model-dependent 
$E_{\max}=\calE_\gamma$ is higher.

Even more stronger upper limit is obtained comparing the {\em numerical  
calculations} of the cascade energy shape with the  Fermi measured spectrum, 
due to its high-energy tail. If the calculated spectrum crosses
the steep high-energy Fermi tail at $E_{\rm cross}$, the calculated flux
above $E_{\rm cross}$ exceeds the observations.  To eliminate this
contradiction  one must lower the total calculated spectrum decreasing
$\omega_{\rm cas}$, and thus we arrive at ``$E_{\max}$-rule'' formulated 
in subsection \ref{sec:omega-numerical} as: ``Increasing $E_{\max}$ in the
calculated spectrum suppresses further $\omega_{\rm cas}$.''

In particular, $E_{\max}$-rule works efficiently in evolutionary 
$(1+z)^m$ models with large $m$ and large $z_{\max}$. Since the flux in
these models is dominated by production at $z_{\max}$, the maximum
energy $E_{\max}$ at $z=0$ becomes $(1+z_{\max})$ times less and
respectively $\omega_{\rm cas}$ is allowed to be higher, as one observes 
in Fig.~\ref{omega:cont}.

We describe now the obtained limits on $\omega_{\rm cas}$ in some
details.  

Fermi LAT presented two kinds of measured  extragalactic fluxes: the total 
Extragalactic Background flux (EGB) and Isotropic Gamma Ray Background 
flux (IGRB). The corresponding energy densities obtained, using  EGB
and IGRB as the upper limits, are $\omega_{\rm cas}^{\rm tot}$ and 
$\omega_{\rm cas}^{\rm iso}$, respectively, the former being always 
larger. Both fluxes, EGR and IGRB, have systematic and statistical 
uncertainties, and are to some extent model dependent (e.g. 
foreground models). To be conservative we use maximal fluxes allowed within 
these uncertainties. 

The both spectra, EGR and IGRB, have steepening, which starts at 
$E_{\rm cut}=250$~GeV and at higher energies become steeper. The method 
of $\omega_{\rm cas}$ calculation is mainly based on this steep high 
energy feature.

The shape of the cascade spectrum is accurately calculated using kinetic 
equations and MC methods. As these calculations show, $E_{\max}$ in 
the cascade spectrum at $z=0$ becomes smaller at larger $z_s$ of 
cascade production. Since flux of EGR/IGRB is $J(E) \propto E^{-3.2}$, 
small $E_{\max}$ results in larger flux $J(E_{\max})$, i.e. in larger  
$\omega_{cas}$. In other words $\omega_{cas}$ is rising with increasing 
$z_s$  as we see indeed in Fig.\ref{omega:point}. 

Fig.~\ref{omega:point} show the case of fixed $z_s$, when strong  
universality holds and thus $\omega_{\rm cas}$ value is unique 
for each $z_s$. As expected $\omega_{\rm cas}^{\rm tot}$ is larger than 
$\omega_{\rm cas}^{\rm iso}$. For small distances $z_s < 0.1$ cascades
are underdeveloped and $\omega_{\rm cas}$ are small:~~
$\omega_{\rm cas}^{\rm iso} \lsim (2-3)\times 10^{-8} eV/cm^3$.
At large $z_s \gsim 1$ energy density is larger:~~
$\omega_{\rm cas}^{\rm iso} \gsim (5-8)\times 10^{-7} eV/cm^3$.
These results depend weakly on models of EBL. 

The strong dependence of $\omega_{\rm cas}$ on redshift $z_s$ 
in Fig.~\ref{omega:point} implies dependence of energy density on 
source distribution over $z$, seen in Fig.~\ref{omega:cont}. One 
may observe there an increase of $\omega_{\rm cas}$ with $z_{\max}$ 
up to  $z_{\max} \sim 1$, with the constant value at larger $z_{\max}$.
This constant  value depends on cosmological evolution $(1+z)^m$. One 
may  summarize the values of $\omega_{\rm cas}^{\rm iso}$ as 
$5\times 10^{-8}eV/cm^3$  in the case of absence of cosmological 
evolution $m=0$ and up to $9\times 10^{-7}eV/cm^3$ in case of 
strong cosmological evolution $m=5$.

The large $\omega_{\rm cas}^{\max}$ allowed in case of strong evolution  
with large $z_{\max}$ is explained by diminishing of $E_{\max}$  
by factor $(1+z)$ in the cascade spectrum at $z=0$.  

The first results of Fermi LAT~\cite{FermiLAT}
demonstrated~\cite{BGKO2011,Ahlers2010,GKS2012} that cascade energy  
density $\omega_{\rm cas} \approx 5.8\times 10^{-7}eV/cm^3$ excludes 
some proton models of UHECR and cosmogenic neutrinos. However, some 
models survived. The new data of Fermi LAT~\cite{Fermi2014}
discovered the steep energy feature in the end of the spectrum
which further constraints the cascade energy density. The new 
limit is model-dependent. For models with strong universality of 
cascade spectrum the limits on $\omega_{\rm cas}$ became stronger
and restrictions on  UHECR became more severe. However, for the models 
with weak universality the restrictions relaxed. In particular,  
the evolutionary models with strong evolution and large $z_{\max}$
the energy density can be larger than $6\times 10^{-7}~eV/cm^3$ ,
i.e. larger than previous limit.

%%%%%%%%%%%%%%%%%%%%%%%%%%%%%%%%%%%%%%%%%%%%%%%%%%%%%%%%%%%%%%%%%%%%%%

{\em Acknowledgments---}%
%%%%%%%%%%%%%%%%%%%%%%%%%%%%%%%%%%%%%%%%%%%%%%%%%%%%%%%%%%%%%%%%%%%%%%
%
Numerical calculations have been performed at the computer cluster of
the  Theoretical Physics Division of the Institute for Nuclear Research
of the Russian Academy of Sciences with support by the Russian Science 
Foundation, grant 14-12-01340.
OK is grateful to GSSI and LNGS for hospitality.\\
%

%%%%%%%%%%%%%%%%%%%%%%%%%%%%%%%%%%%%%%%%%%%%%%%%%%%%%%%%%%%%%%%%%%%%%%%%%%


\begin{thebibliography}{99}
%%%%%%%%%%%%%%%%%%%%%%%%%%%%%%%%%%%%%%%%%%%%%%%%%%%%%%%%%%%%%%%%%%%%%%%%%%%

\bibitem{CMB}
A.~A.~Penzias, R.~W.~Wilson Ap. J. {\bf 142}, 419 (1965).

\bibitem{GZK}
K.~Greisen,
%``End To The Cosmic Ray Spectrum?,''
Phys.\ Rev.\ Lett.\  {\bf 16}, 748 (1966);
G.~T.~Zatsepin and V.~A.~Kuzmin,
%``Upper Limit Of The Spectrum Of Cosmic Rays,''
XCyJETP Lett.\  {\bf 4}, 78 (1966)
[Pisma Zh.\ Eksp.\ Teor.\ Fiz.\  {\bf 4}, 114 (1966)].

\bibitem{BZ1969}
V.~S.~Beresinsky and G.~T.~Zatsepin,
%``Cosmic Rays At Ultrahigh-Energies (Neutrino?),''
Phys.\ Lett.\ B {\bf 28}, 423 (1969);
Sov.\ J.\ Nucl.\ Phys.\ {\bf 11}, 111 (1970).

\bibitem{BS1975}
V. S. Berezinsky and A. Yu. Smirnov,
Astrophys.\ Sp.\ Sci.\ {\bf 32} 461 (1975).
%''Cosmic Neutrinos og Ultra-High Energies and Detection
%possibilities.''

\bibitem{EGRET}
P.~Sreekumar et al., Ap.~J.,{\bf 494}, 523 (1998).
%''EGRET observations of the extragalactic  gamma-ray emission.''

\bibitem{FermiLAT}
A.A.Abdo et al Fermi-LAT Collaboration,
%''The Spectrum of the isotropic diffuse gamma-ray emission derived
%from first-year Fermi Large Area Telescope data.''
Phys Rev Lett. 104. 101101 (2010).

\bibitem{Fermi2014}
M.~Ackermann {\it et al.} [Fermi-LAT Collaboration],
%``The spectrum of isotropic diffuse gamma-ray emission between 100 MeV and 820 GeV,''
Astrophys.\ J.\  {\bf 799}, no. 1, 86 (2015)
[arXiv:1410.3696 [astro-ph.HE]].
%%CITATION = ARXIV:1410.3696;%%

\bibitem{BGKO2011}
V.~Berezinsky, A.~Gazizov, M.~Kachelrie\ss, S.~Ostapchenko,
%''Resricting UHECRs and cosmogenic neutrinos with Fermi-Lat.''
Phys. Lett. B {\bf 695}, 13 (2011), arXiv:1003.1496 .

\bibitem{Ahlers2010}
M.~Ahlers, L.A. Anchordoqui, M.C. Gonzalez-Garcia, F. Halzen, S. Sarkar.
%''GZK neutrinos after the Fermi-LAT diffuse photon flux measurement.''
Astropart. Phys. {\bf 34}, 106 (2010), arXiv:1005.2620

\bibitem{GKS2012}
G.B. Gelmini, O. Kalashev, D.V. Semikoz,
%''Gamma-Ray constraints on maximum cosmogenic neutrino fluxes and
%UHECR source evolution models.''
JCAP {\bf 1201}, 044 (2012),
arXiv:1107.1672

\bibitem{Prilutsky}
O.~F.~Prilutsky, PhD thesis from Moscow Engineering Institute,  1972.

\bibitem{SWW}
A.~W.~Strong, A.~W.~Wolfendale, J.~Wdowczyk, Nature {\bf 241}, 109 (1973).

\bibitem{GouldSchreder}
R.~J.~Gould and D.~Schreder, Phys. ReV. Lett. {\bf 16}, 252 (1966).

\bibitem{Berez1970}
V.~S.~Berezinsky,
% ``Inverse Compton, pair-production and propagation of high energy
% photons and electrons in the Universe.''
Soviet Physics: Nuclear Physics {\bf 11}, 399 (1970) (preprint
P.N. Lebedev Institute of Physics, May 1969).

\bibitem{Felix}
F.~A.~Aharonian, P.~S.~Copi, H.~J.~V\"olk,
%''Very jigh energy cosmic gamma rays from active galactic nuclri:
%Cascading on the cosmic background radiation fields and formation of
%pair halos.
Ap. J. Lett {\bf 423},  L5 (1994).

\bibitem{Kneiske:2003tx}
T.~M.~Kneiske et al.,
%``Implications of Cosmological Gamma-Ray Absorption II. Modification of
%gamma-ray spectra,''
Astron.\ Astrophys.\  {\bf 386}(2002) 1; {\em ibid.},
{\bf 413} (2004) 807.

\bibitem{Kneiske:2010pt}
T.~M.~Kneiske and H.~Dole,
%``A Lower-Limit Flux for the Extragalactic Background Light,''
arXiv:1001.2132 [astro-ph.CO].
%%CITATION = ARXIV:1001.2132;%%
%24 citations counted in INSPIRE as of 14 Jan 2014

\bibitem{NeronovSemikoz}
A.~Neronov, D.~V.~Semikoz,
%''A method of measurement of extragalactic magnetic fields by
%TeV gamma-ray telescopes.''
JETP Letters {\bf 85}, 579 (2007), arXiv:astro-ph/0604607.

\bibitem{Kronberg}
P.~P.~Kronberg,
%''Extragalactic magnetic fields.''
Reports Progress Physics {\bf 57}, 325 (1994).

\bibitem{GrassoRubinstein}
D.~Grasso, H.~R.~Rubinstein,
%''Magnetic fields in early Universe.''
Phys. Rep. {\bf 348}, 163, (2001).

\bibitem{KulsrudZweibel}
R.~M.~Kulsrud, E.~G.~Zweibel,
%''On the origin of cosmic magnetic fields.''
Reports in Progress of Physics {\bf 71(4)} 046,901 (2008). 0707.2783

\bibitem{DurrerNeronov}
R.~Durrer and A.~Neronov,
%``Cosmological Magnetic Fields: Their Generation, Evolution and Observation,''
Astron.\ Astrophys.\ Rev.\  {\bf 21}, 62 (2013)
[arXiv:1303.7121 [astro-ph.CO]].
%%CITATION = ARXIV:1303.7121;%%

\bibitem{LBiermann}
L.~Biermann,
Z.Naturforsch. {\bf 5a}, 65 (1950)

\bibitem{MestelMoss}
L.~Mestel, D.~,L.~Moss,
%''On the Biermann battery process in uniformly rotating, chemically
%inhomogenoous stars.''

\bibitem{MedvedevLoeb}
M.~V.~Medvedev, A.~Loeb,
%''Generation og magnetic fields in the relativistic shock of
%gamma-ray burst sources.''
Ap.J. {\bf 526}, 697 (1999).

\bibitem{MiniatiBell}
F.~Miniati, A.~R.~Bell,
%''Anumerical model of resistive generation of intergalactic magnetic
%field at cosmic dawn.''
Ap.J. {\bf 729}, 73 (2011).

\bibitem{Elyiv2009}
A.~Elyiv, A.~Neronov, D.~V.~Semikoz.
%''Gamma-ray induced cascades and magnetic fields in intergalalactic
%medium''.
arXiv:0903.3649[astro-ph.CO]

\bibitem{NeronovVovk2010}
A.~Neronov, L.~Vovk.
%''Evidence for strong extragalactic magnetic fields from Fermi
%observations of TeV Blazars.''
Science, {\bf 328}, 73 - 75 (2010), arXiv:1006.3504 [astro-phHE].

\bibitem{Taylor:2011bn}
A.~M.~Taylor, I.~Vovk and A.~Neronov,
%``Extragalactic magnetic fields constraints from simultaneous GeV-TeV observations of blazars,''
Astron.\ Astrophys.\  {\bf 529}, A144 (2011)
[arXiv:1101.0932 [astro-ph.HE]].
%%CITATION = ARXIV:1101.0932;%%

\bibitem{TashiroVachaspati}
H.~Tashiro, W.~Chen, F.~Ferrer, T.~Vachaspati, Month. Not. R. Astro.
Soc. {\bf 445}, L41 (2014).

\bibitem{ChenVachaspati}
W.~Chen, B.~D.~Chowdhury, F.~Ferrer, H.~Tashiro, T.~Vachaspati,
arXiv:1412.3171 [astro-ph.CO].

%\bibitem{Long}
%M.~S.~Longair, Mon. Not. {\bf 133}, 421 (1966).

\bibitem{VB1970}
V.~Berezinsky,
%``Inverse Compton effect, pair production and penetration of high energy electrons and photons through the metagalactics,''
Yad.\ Fiz.\  {\bf 11}, 399 (1970).
%%CITATION = YAFIA,11,399;%%
%9 citations counted in INSPIRE as of 16 mar 2015

\bibitem{Blum1970}
G.~R.~Blumenthal and R.~J.~ Gould, Reviews of Modern Physics {\bf 41},
237 (1970).

\bibitem{book}
V.~S.~Berezinsky, S.~ V.~ Bulanov, V.~ A~. Dogiel, V.~ L.~ Ginzburg and
V.~ S.~ Ptuskin, Astrophysics of Cosmic Rays (Elsevier, Amsterdam (1990),
originally in Russian, Nauka (1984).

\bibitem{Kachelriess:2011bi}
M.~Kachelriess, S.~Ostapchenko and R.~Tomas,
%``ELMAG: A Monte Carlo simulation of electromagnetic cascades on the extragalactic background light and in magnetic fields,''
Comput.\ Phys.\ Commun.\  {\bf 183}, 1036 (2012)
[arXiv:1106.5508 [astro-ph.HE]].
%%CITATION = ARXIV:1106.5508;%%
%4 citations counted in INSPIRE as of 16 Dec 2013

\bibitem{BGKO}
V.~Berezinsky, A.~Gazizov, M.~Kachelrie\ss, S.~Ostapchenko,
Phys. Lett. B {\bf 695}, 13 (2011).

\bibitem{Inoue:2012bk}
Y.~Inoue et al.,
% S.~Inoue, M.~A.~R.~Kobayashi, R.~Makiya, Y.~Niino and T.~Totani,
%``Extragalactic Background Light from Hierarchical Galaxy Formation: Gamma-ray Attenuation up to the Epoch of Cosmic Reionization and the First Stars,''
arXiv:1212.1683.
%  [astro-ph.CO].
%%CITATION = ARXIV:1212.1683;%%

\bibitem{okPHD}
O.~K.~Kalashev, PhD thesis, Institute for Nuclear Research RAS, Moscow 2003.\\
O.~E.~Kalashev and E.~Kido,
%``Simulations of Ultra High Energy Cosmic Rays propagation,''
arXiv:1406.0735 [astro-ph.HE].
%%CITATION = ARXIV:1406.0735;%%

\bibitem{Gelmini:2011kg}
G.~B.~Gelmini, O.~Kalashev and D.~V.~Semikoz,
%``Gamma-Ray Constraints on Maximum Cosmogenic Neutrino Fluxes and UHECR Source Evolution Models,''
JCAP {\bf 1201}, 044 (2012).
%   [arXiv:1107.1672 [astro-ph.CO]].
%%CITATION = ARXIV:1107.1672;%%

\bibitem{Lee:1996fp}
S.~Lee,
%``On the propagation of extragalactic high-energy cosmic and gamma-rays,''
Phys.\ Rev.\ D {\bf 58}, 043004 (1998)
[astro-ph/9604098].
%%CITATION = ASTRO-PH/9604098;%%

\bibitem{Yoshida:1998it}
S.~Yoshida, G.~Sigl and S.~-j.~Lee,
%``Extremely high-energy neutrinos, neutrino hot dark matter, and the highest energy cosmic rays,''
Phys.\ Rev.\ Lett.\  {\bf 81}, 5505 (1998)
[hep-ph/9808324].
%%CITATION = HEP-PH/9808324;%%

\bibitem{EBL}
J.~R.~Primack, R.~C.~Gilmore and R.~S.~Somerville,
%``Diffuse Extragalactic Background Radiation,''
AIP Conf.\ Proc.\  {\bf 1085}, 71 (2009);
%   [arXiv:0811.3230 [astro-ph]].
%%CITATION = ARXIV:0811.3230;%%
% \bibitem{Finke:2009xi}
J.~D.~Finke, S.~Razzaque and C.~D.~Dermer,
%``Modeling the Extragalactic Background Light from Stars and Dust,''
Astrophys.\ J.\  {\bf 712}, 238 (2010).
%   [arXiv:0905.1115 [astro-ph.HE]].
%%CITATION = ARXIV:0905.1115;%%

\bibitem{Franceschini:2008tp}
A.~Franceschini, G.~Rodighiero and M.~Vaccari,
%``The extragalactic optical-infrared background radiations, their time evolution and the cosmic photon-photon opacity,''
Astron.\ Astrophys.\  {\bf 487}, 837 (2008)
[arXiv:0805.1841 [astro-ph]].
%%CITATION = ARXIV:0805.1841;%%

\bibitem{Stecker:2005qs}
F.~W.~Stecker, M.~A.~Malkan and S.~T.~Scully,
%``Intergalactic photon spectra from the far ir to the uv lyman limit for 0 < Z < 6 and the optical depth of the universe to high energy gamma-rays,''
Astrophys.\ J.\  {\bf 648}, 774 (2006).
%   [astro-ph/0510449].
%%CITATION = ASTRO-PH/0510449;%%

\bibitem{Stecker:2012ta}
F.~W.~Stecker, M.~A.~Malkan and S.~T.~Scully,
%``A Determination of the Intergalactic Redshift Dependent UV-Optical-NIR Photon Density Using Deep Galaxy Survey Data and the Gamma-ray Opacity of the Universe,''
Astrophys.\ J.\  {\bf 761}, 128 (2012).
%   [arXiv:1205.5168 [astro-ph.HE]].
%%CITATION = ARXIV:1205.5168;%%
%\cite{Scully:2014wpa}
%\bibitem{Scully:2014wpa}
S.~T.~Scully, M.~A.~Malkan and F.~W.~Stecker,
%``An Empirical Determination of the Intergalactic Background Light Using NIR Deep Galaxy Survey Data out to 5 microns and the Gamma-ray Opacity of the Universe,''
arXiv:1401.4435 [astro-ph.HE].
%%CITATION = ARXIV:1401.4435;%%

\bibitem{Essey:2010er}
W.~Essey, O.~Kalashev, A.~Kusenko and J.~F.~Beacom,
%``Role of line-of-sight cosmic ray interactions in forming the spectra of distant blazars in TeV gamma rays and high-energy neutrinos,''
Astrophys.\ J.\  {\bf 731}, 51 (2011).
%   [arXiv:1011.6340 [astro-ph.HE]].
%%CITATION = ARXIV:1011.6340;%%

\bibitem{Abdo:2010kz}
A.~A.~Abdo {\it et al.} % [GBM Collaboration],
%``Fermi Large Area Telescope Constraints on the Gamma-ray Opacity of the Universe,''
Astrophys.\ J.\  {\bf 723}, 1082 (2010).
%   [arXiv:1005.0996 [astro-ph.HE]].
%%CITATION = ARXIV:1005.0996;%%
%59 citations counted in INSPIRE as of 18 Apr 2013

\bibitem{jackson}
For a standard textbook discussion, see J.~D.~Jackson,
{\sl Classical Electrodynamics, 2nd Ed.} (John Wiley \& Sons, New York, 1975).

\bibitem{Liu:2016brs} 
  R.~Y.~Liu, A.~M.~Taylor, X.~Y.~Wang and F.~A.~Aharonian,
  %``Evidence for a Local "Fog" of Sub-Ankle UHECR,''
  arXiv:1603.03223 [astro-ph.HE].
  %%CITATION = ARXIV:1603.03223;%%

\bibitem{Fermi2015}
F.~Acero {\it et al.} [Fermi-LAT Collaboration],
%``Fermi Large Area Telescope Third Source Catalog,''
arXiv:1501.02003 [astro-ph.HE].
%%CITATION = ARXIV:1501.02003;%%

\bibitem{Fermi2015a}
M.~Ackermann {\it et al.} [Fermi-LAT Collaboration],
%``2FHL: The Second Catalog of Hard Fermi-LAT Sources,''
Astrophys.\ J.\ Suppl.\  {\bf 222}, no. 1, 5 (2016)
doi:10.3847/0067-0049/222/1/5
[arXiv:1508.04449 [astro-ph.HE]].
%%CITATION = doi:10.3847/0067-0049/222/1/5;%%

\bibitem{DiMauro2016}
M.~Di Mauro [Fermi-LAT Collaboration],
%``The origin of the Fermi-LAT $\gamma$-ray background,''
arXiv:1601.04323 [astro-ph.HE].
%%CITATION = ARXIV:1601.04323;%%


\end{thebibliography}
\end{document}